\documentclass[pra,twocolumn,a4paper,showpacs,superscriptaddress,floatfix]{revtex4-1}

\usepackage{latexsym}
\usepackage{amsmath}
\usepackage{amsfonts}
\usepackage{graphicx}
\usepackage{amssymb}
\usepackage{dsfont}
\usepackage{color}

\newcommand{\ket}[1]{\ensuremath{|#1\rangle}}
\newcommand{\bra}[1]{\ensuremath{\langle #1 |}}

\newcommand{\be}{\begin{equation}}
\newcommand{\ee}{\end{equation}}
\newcommand{\mc}[1]{\ensuremath{\mathcal{#1}}}
\newcommand{\bc}{\begin{center}}
\newcommand{\ec}{\end{center}}

\newcommand{\mf}[1]{\boldsymbol{#1}}

\newcommand{\ve}{\varepsilon}

\newcommand{\vro}{\varrho}

\newcommand{\rf}{\vro_{\text{B}}}
\newcommand{\hf}{\mc{H}_{\text{FE}}}
\newcommand{\rd}{\vro_{\text{D}}}
\newcommand{\hc}{\mc{H}_{\text{tot}}}

\newcommand{\Si}[2]{S_{#1}^{(#2)}}

\newcommand{\mean}[1]{\ensuremath{ \langle #1  \rangle}}
\newcommand{\meanX}[1]{\| #1  \|}

\begin{document}

\title{Master equation approach for interacting slow- and stationary-light  polaritons }

\author{M. \surname{Kiffner}}
\affiliation{Technische Universit\"at M\"unchen, Physik-Department I, 
James-Franck-Stra{\ss}e, 85748 Garching, Germany}

\author{M.~J. \surname{Hartmann}}
\affiliation{Technische Universit\"at M\"unchen, Physik-Department I, 
James-Franck-Stra{\ss}e, 85748 Garching, Germany}

\begin{abstract}
A master equation approach for the description of dark-state polaritons in 
coherently driven atomic media is presented. This technique provides a  description of 
light-matter interactions under conditions of electromagnetically induced 
transparency (EIT) that is well suited for the treatment of polariton losses. 
The master equation approach allows us to 
describe general polariton-polariton interactions that may be conservative, dissipative 
or a mixture of both. In particular, it enables us to study 
dissipation-induced correlations  as a means for the creation of strongly correlated 
polariton systems. 
Our technique reveals a loss mechanism for 
stationary-light polaritons that has not been discussed so far. 
We find that polariton losses in level configurations with non-degenerate 
ground states can be a multiple of those in level schemes with degenerate ground states. 
\end{abstract}

\maketitle

\section{INTRODUCTION \label{intro}}
Photons are ideal carriers for quantum information over long distances.  
This is due to the large propagation speed of light and the fact that photons 
in free space do not interact with each other.  On the other hand,  the generation 
of highly entangled light fields and the realization of photon gates requires 
strong photon-photon interactions~\cite{nielsen:00}. 
Nonlinear media can mediate an effective interaction between photons, but the strength 
of this induced coupling is usually weak. Thus the realization of  strong photon-photon  
interactions is a major challenge in quantum information science. 
Similarly, strong photon-photon interactions are  a key requirement for quantum optical 
implementations of highly correlated  many-body systems~\cite{bloch:08}. A substantial 
research effort
~\cite{HBP06,HBP08,hartmann:08,RF07,ASB06,GTI+09,hafezi:09a,hafezi:09,carusotto:09,fleischhauer:08,chang:08,kiffner:10} 
is currently devoted to these systems where combined excitations of light and matter, 
i.e. polaritons, reproduce the dynamics of bosons with tunable mass and different 
interaction types. Several effects in correlated  many-body systems  were considered, 
including  the realization of   Bose-Hubbard models~\cite{HBP06,hartmann:08,HBP08,RF07,ASB06}, 
quantum transport~\cite{hafezi:09a,hafezi:09}, nonlinear effects in  driven dissipative 
systems~\cite{GTI+09,hartmann:10}, Bose-Einstein condensation~\cite{fleischhauer:08}, and 
the realization of a Tonks-Girardeau gas~\cite{chang:08,carusotto:09,kiffner:10} 
of polaritons. 

Dark-state polaritons~\cite{fleischhauer:00,fleischhauer:02,lukin:03} represent   
bosonic quasi-particles that arise in light-matter interactions under conditions of electromagnetically 
induced transparency (EIT)~\cite{fleischhauer:05}. 
The generic EIT scheme consists of a gas comprised of three-level atoms in $\Lambda$  
configuration that are driven by  a strong control field and a weak probe field on 
separate transitions. 
EIT gives rise to a multitude of intriguing effects like  the slowing and stopping of 
light~\cite{hau:99,kash:99,budker:99}, the   coherent storage and retrieval of 
light~\cite{hau:01,phillips:01, fleischhauer:00,fleischhauer:02,dey:03} 
and stationary light
~\cite{andre:02,bajcsy:03,zimmer:06,zimmer:08,moiseev:05,moiseev:06,fleischhauer:08,lin:09,nikoghosyan:09}.  
Of particular relevance for quantum optical realizations of strongly correlated 
many-body  systems is stationary light where the polaritons experience a quadratic 
dispersion relation like bosons in free space.  
Moreover, EIT in atomic four-level systems gives rise to a strongly enhanced Kerr nonlinearity 
for the probe fields~\cite{schmidt:96,ISWD97,harris:98,harris:99,kang:03,braje:04,hartmann:07}. 
This effect results in a two-particle contact interaction between the polaritons that 
can be conservative~\cite{andre:05,chang:08} or dissipative~\cite{kiffner:10}. 
%

\begin{figure}[b!]
\includegraphics[width=8cm]{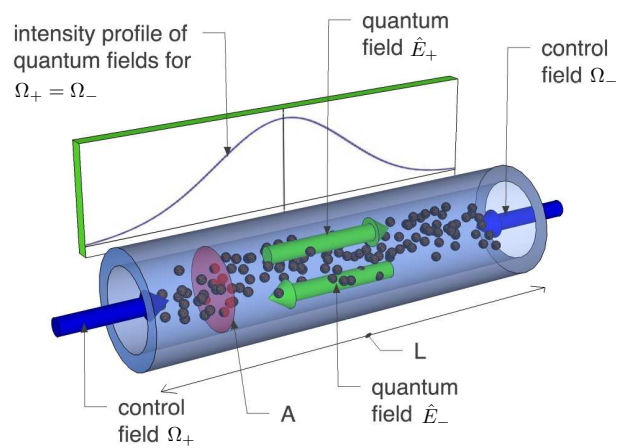}
\caption{\label{fig1} 
(Color online)   Considered setup of 
$N$ atoms   confined to an interaction volume of length $L$ and transverse area $A$. 
$\Omega_{\pm}$ are the Rabi frequencies of the classical control fields, and 
$\hat{E}_{\pm}$  are the quantum probe fields. A typical intensity profile of 
the quantum fields is shown for the case of stationary light where the 
control fields have the same intensity ($\Omega_+=\Omega_-$). 
}
\end{figure}
\begin{figure}[t!]
\includegraphics[width=6.5cm]{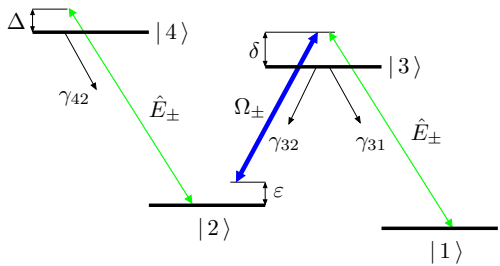}
\caption{\label{fig2} 
(Color online)  Atomic level scheme. $\gamma_{ij}$ is the full decay rate 
on the $\ket{i}\leftrightarrow\ket{j}$ transition, $\delta$ and $\Delta$ 
label the detuning of the probe fields with states $\ket{3}$ and $\ket{4}$, 
respectively, and $\ve$ is the two-photon detuning. 
}
\end{figure}

%
Experimental setups that show great promise for the realization of 
strongly correlated polariton systems  are arrays of coupled microcavities 
doped with  emitters~\cite{aoki:06,wallraff:04,leib:10,hennessy:07,trupke:07} or optical fibers that couple to 
atoms~\cite{bajcsy:09,vetsch:09}. In general, strong light-matter interactions 
require the confinement of light to small interaction volumes. 
Here we consider the experimental setup~\cite{bajcsy:09} shown in Fig.~\ref{fig1}, 
where photons and atoms are simultaneously confined to the hollow core of a 
photonic-crystal fiber, and the level scheme of each atom is shown in Fig.~\ref{fig2}. 
Since the light-guiding core of the optical fiber is of the same order of magnitude 
as the optical wavelength, the fiber represents a one-dimensional waveguide 
for the optical fields. Note that a second potential realization is comprised of 
the experimental setup in~\cite{vetsch:09}, where multi-color evanescent 
light fields surrounding an optical nanofiber couple to atoms trapped in 
an optical lattice. 

Existing descriptions~\cite{fleischhauer:00,fleischhauer:02} of dark-state 
polaritons in EIT systems are based on a Heisenberg-Langevin approach for the 
polariton field operator.  Here we present a   different approach and derive a 
master equation for the reduced density operator of  dark-state polaritons. 
The master equation technique  facilitates the treatment of 
polariton losses and allows one to 
account for general polariton-polariton interactions 
that may be conservative, dissipative or a mixture of both. This is an important 
achievement since it opens up the possibility to study dissipation-induced 
correlations~\cite{syassen:08,duerr:09} in polariton systems. Dissipative 
polariton-polariton interactions are  promising  in the quest for highly correlated 
systems since they can be considerably stronger than their conservative counterparts~\cite{kiffner:10}. 
Second, our method reveals an additional  loss term for  stationary-light polaritons 
whose importance depends on the structure of the atomic level scheme and that was 
not discussed in  the literature yet. 
%

%
This paper is organized as follows. In  Sec.~\ref{description} 
we set up a master equation for the atoms interacting with
the quantized probe and classical control fields inside the 1D waveguide, 
see Figs.~\ref{fig1} and~\ref{fig2}. 
We then transfer the original master equation into a master equation 
solely for dark-state polaritons. This process is detailed in Sec.~\ref{Smeq} 
and consists of four steps. First, we show that the entire problem can 
be described in terms of bosonic quasi-particles if the number of atoms 
is much larger than the number of probe field photons, see Sec.~\ref{excitations}. 
The concept of dark-state polaritons is introduced in Sec.~\ref{DSP}, 
and the formulation of the original master equation of Sec.~\ref{description} 
in terms of dark-state polaritons and all other excitations is presented in Sec.~\ref{model}.  
In the final step of the derivation we  trace out all excitations except for 
the dark-state polaritons, see  Sec.~\ref{elemination}. 
The master equation for dark-state polaritons under conditions of stationary 
light and for arbitrary (conservative or dissipative) polariton-polariton interactions 
is presented in Sec.~\ref{resmeq}.  Here we summarize all conditions that grant 
the validity of our approach. The special case where the level scheme in Fig.~\ref{fig2} 
is reduced to the $\Lambda$ subsystem  
is discussed in Sec.~\ref{statlight}. We compare the predictions of our master equation 
to the results of a numerical integration of Maxwell-Bloch equations and find excellent 
agreement. 
The full master equation including the most general form of polariton-polariton 
interactions is covered in Sec.~\ref{nonlin}, 
and the mapping to the dissipative Lieb-Liniger model is outlined in Sec.~\ref{dll}. 
Finally, in Sec.~\ref{other} we compare the advantages and disadvantages 
between the  atomic level schemes in Figs.~\ref{fig2} and~\ref{fig5} that both give 
rise to the same master equation for dark-state polaritons. 
\section{DESCRIPTION OF THE SYSTEM \label{description}}
We start with a more detailed description of 
our one-dimensional model shown in Figs.~\ref{fig1} and~\ref{fig2}. 
Each of the $N$ atoms interacts with  control and probe  fields denoted by $\Omega_{\pm}$ and 
$\hat{E}_{\pm}$, respectively. 
The control fields of frequency $\omega_c$ are treated  classically 
and $\Omega_+$ ($\Omega_-$) labels  the Rabi frequency 
of the control field propagating in the positive (negative) $z$ direction.  
In addition, we assume that the control fields are spatially homogeneous and 
that the Rabi frequencies $\Omega_{\pm}$ are real. 
The probe fields $\hat{E}_+$ and $\hat{E}_-$ are quantum fields 
that propagate in the positive and negative $z$ direction, respectively. They  
are defined as 
\be
\hat{E}_{\pm}(z) = \sum_{k} a_{\pm k_c + k }e^{ i (\pm k_c +k)  z}, 
\label{probefield}
\ee
where $a_{\pm k_c + k}$  are photon annihilation operators of a mode 
with frequency $\omega_{\pm k_c +k}$ and
$k_c$ ($-k_c$) is the wave number of the control field $\Omega_+$ ($\Omega_-$). 
We assume  that the wave numbers $k$ satisfy $|k|\ll k_c$ which implies 
that the envelope of the quantum fields varies slowly on a lengthscale 
defined by the wavelength of the optical fields.  

We model the  time evolution of the atoms and the quantized probe fields  by a 
master equation~\cite{breuer:os} for their density operator $\vro$, 
\be
\dot{\vro}= - \frac{i}{\hbar}[ H ,\vro] + \mc{L}_{\gamma}\vro, \label{meq0}
\ee
where the  system Hamiltonian 
$H  = H_0 + H_{\Lambda} + H_{\text{NL}}$ is comprised of three parts. 
$H_0$ describes the free time evolution of the atoms and the probe fields, 
and $H_{\Lambda}$ accounts for the interaction   of the  probe and  control fields with 
the $\Lambda$-subsystem formed by states $\ket{1}$, $\ket{2}$ and $\ket{3}$. 
On the other hand, $H_{\text{NL}}$ accounts for the  coupling of the probe fields 
to the $\ket{2}\leftrightarrow\ket{4}$ transition and results in a nonlinear coupling 
between probe field photons~\cite{schmidt:96,ISWD97,harris:98,harris:99,kang:03,braje:04,hartmann:07}. 
In a rotating frame that removes the time-dependence of the classical laser 
fields, $H_0$,   $H_{\Lambda}$ and $H_{\text{NL}}$ are given by 
\begin{align}
 H_0 = & -\hbar \sum_{k}(\omega_p-\omega_{k_c+k}) a^{\dagger}_{k_c+k} a_{k_c+k} \label{H0} \\ 
& -\hbar \sum_{k}(\omega_p-\omega_{-k_c+k}) a^{\dagger}_{-k_c+k } a_{-k_c+k}  \notag \\
& -\hbar 
\sum_{\mu=1}^N\left[ \varepsilon A_{22}^{(\mu)} + \delta A_{33}^{(\mu)} + 
(\Delta + \varepsilon)A_{44}^{(\mu)}\right] ,\notag
\end{align}
\begin{align}
H_{\Lambda} = & - \hbar\sum_{\mu=1}^{N}\Big\{ \Si{32}{\mu} 
\left[ \Omega_+ e^{i k_c z_{\mu}} + \Omega_- e^{-i k_c z_{\mu}}\right] \label{hlambda} \\
& \hspace*{0.5cm} + g_1 
\Si{31}{\mu}  \left[ \hat{E}_+(z_{\mu}) +\hat{E}_-(z_{\mu}) \right] \Big\}
 +\text{h.c.}  ,   \notag  
\end{align}
\begin{align}
H_{\text{NL}} =  &  - \hbar g_2 \sum_{\mu=1}^{N}
 \Si{42}{\mu}  \left[ \hat{E}_+(z_{\mu}) +\hat{E}_-(z_{\mu}) \right]
 +\text{h.c.}  \,. \label{hnl}
\end{align}
Here $A_{ii}^{\mu}$ and $\Si{ij}{\mu}$ are projection and transition operators 
of atom $\mu$ at position $z_{\mu}$, respectively,
\begin{align}
& A_{ii}^{\mu} = \ket{i_{\mu}}\bra{i_{\mu}}, &&
\Si{ij}{\mu} = \ket{i_{\mu}}\bra{j_{\mu}} \quad (i\not=j), 
\end{align}
$\omega_p$ is the central frequency of the probe pulse, 
and $g_1$ ($g_2$) is the single-photon Rabi frequency on the 
$\ket{3}\leftrightarrow\ket{1}$ ($\ket{4}\leftrightarrow\ket{2}$) 
transition.  
The  detuning of the probe field with respect to the transition 
$\ket{3}\leftrightarrow\ket{1}$ ($\ket{4}\leftrightarrow\ket{2}$)  
is labeled by  $\delta$ ($\Delta$), 
and $\ve$ is the two-photon detuning, 
\begin{align}
\delta  = \omega_p - \omega_{31},\quad
\Delta =  \omega_p - \omega_{42}, \quad
\ve  = \omega_p-\omega_c -\omega_2. \label{dets}
\end{align}
Here the energy of level $\ket{i}$ is 
$\hbar \omega_i$ (we set  $\omega_1=0$) and transition frequencies are denoted 
by $\omega_{ij}=\omega_i-\omega_j$.  
The term $\mc{L}_{\gamma}\vro$ in Eq.~(\ref{meq0}) 
accounts for spontaneous emission from states $\ket{3}$ and $\ket{4}$, 
\begin{align}
\mc{L}_{\gamma}\vro = & -\frac{\gamma_{31}}{2}\sum\limits_{\mu=1}^{N}
\left( A_{33}^{(\mu)} \vro + \vro A_{33}^{(\mu)}  - 2 S_{13}^{(\mu)} \vro S_{31}^{(\mu)} \right) \label{decay0} \\
& -\frac{\gamma_{32}}{2}\sum\limits_{\mu=1}^{N}
\left( A_{33}^{(\mu)} \vro + \vro A_{33}^{(\mu)}  - 2 S_{23}^{(\mu)} \vro S_{32}^{(\mu)} \right) \notag \\
& -\frac{\gamma_{42}}{2}\sum\limits_{\mu=1}^{N}
\left( A_{44}^{(\mu)} \vro + \vro A_{44}^{(\mu)}  -2 S_{24}^{(\mu)} \vro S_{42}^{(\mu)} \right) ,\notag 
\end{align}
where $\gamma_{ij}$  is the full decay rate on the transition $\ket{i}\leftrightarrow\ket{j}$ 
(see Fig.~\ref{fig1}). 
Finally, we introduce the parameter 
\be
\Delta\omega = \omega_p-\omega_c = \omega_2 +\ve \label{do}
\ee
which describes the frequency difference between the probe and control fields. 
Note that  $\Delta\omega$ is practically equal 
to the frequency splitting $\omega_2$ between the ground states $\ket{1}$ and $\ket{2}$ 
if the  two-photon detuning $\ve$ is small. 
Next we outline the approach we developed to reduce the master equation~(\ref{meq0}) 
for the atoms and quantized probe fields into a master equation solely for dark-state 
polaritons~\cite{fleischhauer:00}, formed by collective excitations of photons and atoms.
\section{MASTER EQUATION FOR DARK-STATE POLARITONS: DERIVATION \label{Smeq}}
Here we show that the master equation~(\ref{meq0}) can be simplified considerably 
if we assume that almost all atoms are in state $\ket{1}$ and 
that the total number of photons is much smaller than the number of atoms $N$. 
This assumption allows us to study the system dynamics entirely in terms of independent 
bosonic quasi-particle excitations, see Sec.~\ref{excitations}. 
A second simplification is made possible by 
the concept of dark-state polaritons~\cite{fleischhauer:00,fleischhauer:02} introduced in Sec.~\ref{DSP}.  
Dark-state polaritons are bosonic quasi-particles that decay only indirectly via 
the coupling to other bosonic modes that are termed bath excitations. 
In the so-called slow-light regime,  this coupling  is  much slower than 
the decay of the bath excitations which is 
of the order of  the decay rates of the excited states $\ket{3}$ and $\ket{4}$.  
The existence of these two different time scales enables us to derive a 
Markovian master equation for the reduced density operator of the dark-state polaritons 
as outlined in  Sec.~\ref{elemination}. 
Throughout this Section, all technical details and lengthy definitions 
are moved to the Appendix.
\subsection{BOSONIZATION \label{excitations}}
Here we show that the system described in Sec.~\ref{description} can 
be mapped to a much simpler system if almost all atoms are in state $\ket{1}$ and 
if the total number of photons is much smaller than the number of atoms $N$. 
We begin with the description of a simple system 
that consists of $M$ bosonic modes. The annihilation operator  of mode $i$  
is given by  $O_i$ ($i \in \{1,\ldots,M\}$) and 
$\mc{O}=\{O_1,\ldots,O_M\}$ denotes the set of all operators that 
obey the commutation relations  
\be
\left[ \hat{O}_i,\hat{O}_j^{\dagger}\right] = \delta(i,j),\quad 
\left[ \hat{O}_i,\hat{O}_j\right] = 0 . \label{bcr}
\ee
If $\ket{0}$ is the vacuum state of the system, it follows that
\be 
\ket{\{n_1,\ldots,n_M\}} = \prod_{i=1}^{M}\frac{1}{\sqrt{n_i!}}\big(\hat{O}_i^{\dagger}\big)^{n_i}\ket{0} 
\label{states}
\ee
is a normalized state with  $n_i$ excitations in mode $i$ and a total 
number of $\sum_{i=1}^M n_i$ excitations. 
Furthermore, we note that the total state space of   $M$ bosonic modes is the 
tensor product of the state spaces $\mc{H}_i$ associated with the individual modes, 
\be
\mc{H}_{\text{osc}} = \mc{H}_1 \otimes\mc{H}_2\otimes\ldots\otimes\mc{H}_M . 
\label{stateSp}
\ee

Next we show how the system of Sec.~\ref{description} can be mapped to 
this simple model outlined in Eqs.~(\ref{bcr})-(\ref{stateSp}). 
First we define a vacuum state 
where all probe field modes are empty and all atoms are in state $\ket{1}$, 
\be
\ket{0}=\ket{\{0\}_{\text{phot}};1_1,\ldots,1_N}. \label{vac}
\ee
Second, we define the following  operators ($m\in \mathds{Z}$)
\begin{align}
& A_k=a_{k_c + k}\sin\varphi   + a_{-k_c + k}\cos\varphi , \label{ak} \\
& D_k=a_{k_c + k} \cos\varphi - a_{-k_c + k} \sin\varphi , \label{dk} \\
& X_k(m)= \frac{1}{\sqrt{N}}\sum_{\mu=1}^{N} \Si{12}{\mu} e^{-i(m k_c+ k) z_{\mu}}, \label{xk} \\
& H_k(m) = \frac{1}{\sqrt{N}}\sum_{\mu=1}^{N} \Si{13}{\mu} e^{-i(m k_c+ k) z_{\mu}}, \label{hk} \\
& I_k(m) = \frac{1}{\sqrt{N}}\sum_{\mu=1}^{N} \Si{14}{\mu} e^{-i(m k_c+ k) z_{\mu}},  \label{ik}
\end{align}
where $k=n 2\pi/L$ ($n \in \mathds{Z}$) and 
$A_k$ ($D_k$) is a sum (difference) of two counter-propagating probe field modes. 
Since $a_{\pm k_c+ k}$ are photon annihilation operators, it follows that $A_k$ and $D_k$ 
obey bosonic commutation relations. The angle $\varphi$ depends on the relative strength of 
the Rabi frequencies $\Omega_+$ and $\Omega_-$ and is defined 
by~\cite{zimmer:06,zimmer:08}  
\begin{align}
& \sin \varphi =  \frac{\Omega_+}{\sqrt{\Omega_+^2 + \Omega_-^2}} , \quad
 \cos \varphi =  \frac{\Omega_-}{\sqrt{\Omega_+^2 + \Omega_-^2}} .
\end{align}
The operator $X_k(m)$ describes a collective spin coherence that is slowly oscillating for $m=0$ 
and fast oscillating for $m\not= 0$.
The operators $H_k^{\dagger}(m)$ and $I_k^{\dagger}(m)$ create an excitation 
in the excited states $\ket{3}$ and $\ket{4}$, respectively. 
Next we show that the operators defined in Eqs.~(\ref{xk})-(\ref{ik}) 
obey  bosonic commutation relations for all wave numbers $k$ and all $m\in \mathds{Z}$ 
if almost all atoms are in state $\ket{1}$. 
As an example, we discuss the commutation relations for $X_k(m)$. 
Within a manifold with fixed $m$, we find 
\begin{align}
\left[ X_k(m),X_p^{\dagger}(m)\right] & 
= \frac{1}{N} \sum\limits_{\mu=1}^{N} \left(A_{11}^{(\mu)} - A_{22}^{(\mu)}\right)
e^{i(p-k)z_{\mu}}  \label{bc} \\ 
& \approx \frac{1}{N} \sum\limits_{\mu=1}^{N} e^{i(p-k)z_{\mu}} \\
& \approx \frac{1}{L} \int\limits_{0}^{L}dz  e^{i(p-k)z}=\delta(k,p)\,, \notag 
\end{align}
where we set $A_{11}^{(\mu)}\approx\mathds{1}$ and $A_{22}^{(\mu)}\approx 0$  
since almost all atoms are in state $\ket{1}$. 
Furthermore,  we employed that the mean distance between atoms is much smaller 
than $1/|k|$ for all relevant wavenumbers $k$ contributing to  the slowly varying 
envelopes of the control fields. 
In the case  $m\not=n$, we find 
\begin{align}
\left[ X_k(m), X_p^{\dagger}(n)\right] 
& \approx \frac{1}{N} \sum\limits_{\mu=1}^{N} e^{i[(p-k)+(n-m)k_c]z_{\mu}}. 
\label{bc2}
\end{align}
In contrast to Eq.~(\ref{bc}), the sum cannot be converted into an integral since 
the mean spacing between the atoms is   much larger than $1/k_c$ for realistic 
densities, where 
$k_c$ is the wavenumber of an optical transition. However, the sum in Eq.~(\ref{bc2}) 
represents the average of random numbers on the unit circle in the complex plain, which is 
zero in the limit $N\rightarrow \infty$. We can thus set 
$[ X_k(m), X_p^{\dagger}(n) ]\approx 0$ for $m\not=n$. 
Since $[X_k(m),X_p(n)] =0$, it follows that the operators $X_k(m)$ 
obey bosonic commutation relations if almost all atoms are in state $\ket{1}$, 
and corrections  scale with $1/N$. 
The same result is found for $H_k(m)$ and $I_k(m)$. Furthermore, we point out that 
$X_k(m)$, $H_k(m)$ and $I_k(m)$ describe independent excitations up to corrections that 
scale with $1/N$, i.e., $[ X_k(m), H_p^{\dagger}(n) ]\approx 0$, 
$[ X_k(m), I_p^{\dagger}(n) ]\approx 0$ and $[ I_k(m), H_p^{\dagger}(n) ]\approx 0$. 
In summary, we can introduce  the set of bosonic operators 
\be
\mc{O} = 
\{A_k , D_k , X_k(m) , H_k(m) , I_k(m)\}_{\genfrac{}{}{0pt}{2}{k \hspace*{0.4cm}}{m\in\mathds{Z}}} 
\label{operators}
\ee
if almost all atoms are in state $\ket{1}$. 
This condition can be met if the total state space $\hc$ of the 
system  is restricted to the subspace  
\be
\hf = \text{Span}\left\{\ket{\{n_1,\ldots,n_M\}} 
\;\big|\; \sum\limits_{i=1}^M n_i \ll N
 \right\} \label{hf}
\ee 
that is spanned by states with much less excitations than number of atoms $N$. 
From a physical point of view, the system dynamics will be restricted to 
this subspace if the number of photons is much smaller than the number of 
atoms, and if initially almost all atoms are in state $\ket{1}$. 
In Appendix~\ref{representation}, we 
show that the Hamiltonian $H$ and the decay term $\mc{L}_{\gamma}\vro$  
can be expressed entirely in terms of  the 
bosonic operators in Eq.~(\ref{operators}) if the state space is restricted 
to $\hf$. We denote the density operator in $\hf$ by $\tilde{\vro}$, 
and the master equation~(\ref{meq0}) in $\hf$ can be written as 
\be 
\dot{\tilde{\vro}} = -\frac{i}{\hbar}[\tilde{H},\tilde{\vro}] 
+ \tilde{\mc{L}}_{\gamma} \tilde{\vro} ,  \label{meq1}
\ee
where 
$\tilde{H}  = \tilde{H}_0 + \tilde{H}_{\Lambda} + \tilde{H}_{\text{NL}}$.  
The operators $\tilde{H}_0$, $\tilde{H}_{\Lambda}$, $\tilde{H}_{\text{NL}}$  
and $\tilde{\mc{L}}_{\gamma}\tilde{\vro}$ are defined in Eqs.~(\ref{h0fe}),~(\ref{hlambdafe}),~(\ref{hnlfe}) 
and~(\ref{decay1}), respectively. 
These operators approximate  their counterparts without tilde in the subspace $\hf$ and 
are comprised of the  bosonic operators in Eq.~(\ref{operators}). 

\subsection{DARK-STATE POLARITONS \label{DSP}}
The interaction Hamiltonian $H_{\Lambda}$ of the $\Lambda$-subsystem 
has the important property that a certain class of its eigenstates are 
so-called dark states $\ket{D}$. These states  are called dark since 
they do not contain a  contribution of the excited state $\ket{3}$ 
and are thus immune against spontaneous emission. 
A simple example for a   dark state is given by 
\be
\ket{D} = \psi_k^{\dagger}\ket{0}, \label{dss}
\ee
where the unique definition of the  operator $\psi_k$ is~\cite{zimmer:08}
\begin{align}
\psi_k = A_k \cos\theta  - X_k \sin\theta.  
\label{dsp}
\end{align}
In this equation, $A_k$ 
is a superposition of two counter-propagating probe field modes, and 
$X_k=X_k(0)$ 
describes a slowly varying collective spin coherence. 
The mixing angle $\theta$ determines the weight of the 
photonic ($A_k$) and atomic ($X_k$) components contributing to $\psi_k$ 
and is defined as~\cite{fleischhauer:00,fleischhauer:02,andre:02,zimmer:06,zimmer:08} 
\begin{align}
& \sin \theta =  \frac{\sqrt{N} g_1}{\Omega_0} , \;
 \cos \theta =  \frac{\sqrt{\Omega_+^2 + \Omega_-^2}}{\Omega_0}, \;
\\
& 
\Omega_0 = \sqrt{N g_1^2 + \Omega_+^2 + \Omega_-^2 }.  \label{scO}
\end{align}
A short  calculation shows that $\ket{D}$ in Eq.~(\ref{dss}) is  an  eigenstate 
of  $H_{\Lambda}$ in Eq.~(\ref{hlambda}) with eigenvalue zero, i.e., $H_{\Lambda}\ket{D}=0$. 
Furthermore,  Eq.~(\ref{dsp}) implies that $\ket{D}$ does not contain a contribution of 
the excited state $\ket{3}$. It follows that state $\ket{D}$  
is indeed  a dark state of the interaction Hamiltonian $H_{\Lambda}$. 
Note, however, that $\ket{D}$ is not an eigenstate of  the remaining 
parts $H_0$ and $H_{\text{NL}}$ of the full Hamiltonian, and   these terms 
give rise to a non-trivial time evolution of the dark-state polaritons. 

The results of Sec.~\ref{excitations} and Eq.~(\ref{dsp}) imply that 
the operators $\psi_k$ obey bosonic commutation relations in $\hf$,
\begin{align}
\left[\psi_k,\psi_p^{\dagger}\right] \approx \delta(k,p),
\label{bcd}
\end{align}
and the  quasi-particles associated with these excitations 
are termed dark-state polaritons. 
It follows from  Eq.~(\ref{bcd}) that 
\be
\ket{D} =
 \prod_{k}\frac{1}{\sqrt{n_k!}}\big(\psi_k^{\dagger}\big)^{n_k}\ket{0} 
\label{Dgen}
\ee
represents a normalized dark state with $\sum_{k}n_k$ excitations. 
We emphasize  that the photonic part $A_k$ of $\psi_k$ 
contains a pair of counter-propagating 
probe field modes that are grouped around the wavenumbers  
of the control fields $\pm k_c$ 
rather than the mean  wavenumbers $\pm k_p$ of the probe fields, see Eq.~(\ref{ak}). 
Note  that 
the states in Eq.~(\ref{Dgen}) would  not be true dark states 
if the probe field modes in $A_k$ were grouped around $\pm k_p$ and if both $\Omega_+$ 
and $\Omega_-$ were different from zero.

\subsection{BOSONIZATION WITH DARK-STATE POLARITONS \label{model}}
We have shown in Sec.~\ref{excitations} that the master equation~(\ref{meq0}) can 
be formulated in terms of bosonic modes if the system dynamics is restricted to the 
subspace $\hf$. 
Here we restate this model in terms of long-lived dark-state polaritons introduced 
in Sec.~\ref{DSP}.
With the definition of the bright-state polariton~\cite{fleischhauer:08}  
\be 
\phi_k = A_k \sin\theta + X_k \cos\theta  \label{bp}
\ee
and the inverse relations of Eqs.~(\ref{dsp}) and~(\ref{bp}), 
we replace the operators $A_k$ and $X_k(0)$ in $\mc{O}$ [see Eq.~(\ref{operators})] 
by $\psi_k$ and  $\phi_k$. 
This allows us to write the  subspace $\hf$   in Eq.~(\ref{hf}) 
as  the tensor product of the state space $\mc{H}_{\text{S}}$ 
of dark-state polaritons  
and the state space $\mc{H}_{\text{B}}$ of all other modes termed bath excitations, 
\be 
\hf = \mc{H}_{\text{S}}\otimes \mc{H}_{\text{B}}\,. 
\ee
The partition of all bosonic modes into dark-state polaritons and 
bath excitations is motivated by our aim to derive a master equation for the long-lived 
dark-state polaritons only, see Sec.~\ref{elemination}.

In the following, we assume that the Rabi frequencies of the control fields are 
identical  and set 
\be
\Omega_c = \Omega_+ = \Omega_- . \label{equalR}
\ee
With this choice, $H_{\Lambda}$ gives rise to the stationary light phenomenon~\cite{andre:02,bajcsy:03} 
that allows us to trap the probe field inside the medium. 
Note that any other  choice  of the Rabi frequencies $\Omega_{\pm}$ can be treated within the 
formalism introduced here, and in these cases the calculation follows exactly the same route as 
detailed below. 
If  the operators $A_k$ and $X_k(0)$ in the master equation~(\ref{meq1}) are replaced 
by $\psi_k$ and  $\phi_k$,  we obtain (see Appendix~\ref{definitions})
\be 
\dot{\tilde{\vro}} = -\frac{i}{\hbar}[H_{\text{S}},\tilde{\vro}] -\frac{i}{\hbar}[V,\tilde{\vro}] 
+ \mc{L}_{\text{B}} \vro ,  \label{meq2}
\ee
where 
\be
H_{\text{S}} = -\hbar(\sin\theta \ve + \cos^2\theta)\sum\limits_k \psi_k^{\dagger}\psi_k 
\ee
describes the free time evolution of the dark-state polaritons. 
Here we choose a small two-photon detuning
\be 
\ve = -\cot^2\theta \Delta\omega  \label{twophot}
\ee
such that $H_{\text{S}}=0$. 
The dynamics of the  bath excitations is governed by the Liouvillian
\begin{align}
\mc{L}_{\text{B}} \tilde{\vro} = -\frac{i}{\hbar}[H_{\text{B}},\tilde{\vro}]
+ \mc{L}_{\gamma}^{(\text{B})}\tilde{\vro}  , \label{lr}
\end{align}
where $H_{\text{B}}$ accounts for the unitary time evolution of the bath modes 
and the decay of bath excitations is determined by $\mc{L}_{\gamma}^{(\text{B})}\tilde{\vro}$. 
The interaction between dark-state polaritons and 
other excitations in $\mc{H}_{\text{B}}$ is described by the interaction Hamiltonian $V$. 
The definitions of $H_{\text{B}}$, $\mc{L}_{\gamma}^{(\text{B})}\tilde{\vro}$ 
and $V$ are provided in Appendix~\ref{definitions}. 

The  decay term $\mc{L}_{\gamma}^{(\text{B})}\tilde{\vro}$ results in 
a finite lifetime of the bath excitations that is of the order of 
the lifetimes of the excited states $\ket{3}$ and $\ket{4}$. On the other hand, 
the dark-state polaritons decay only indirectly via the coupling to bath 
excitations mediated by $V$. In the slow light limit, this coupling is much 
slower than the decay of the bath excitations. This existence of two 
different time scales opens up the possibility to derive a Markovian 
master equation for the dark-state polaritons alone, and this 
procedure is outlined in the next Section~\ref{elemination}. 
\subsection{ELIMINATION OF THE BATH \label{elemination}}
In the previous  Sections~\ref{excitations}-\ref{model} we achieved to transform  
the initial master equation~(\ref{meq0}) within the subspace $\hf$  into 
a master equation for long-lived dark-state polaritons and fast-decaying bath excitations. 
Here we are especially interested in the quantum state $\rd$ of the dark-state polaritons 
that is obtained from $\tilde{\vro}$ by a partial trace over 
all excitations except for the dark state polaritons. 
We derive the corresponding  master equation for $\rd$   from Eq.~(\ref{meq2})   
via projection operator techniques~\cite{breuer:os} and assume 
that the initial state of the system factorizes into a product of the 
initial polariton state $\rd$ and the vacuum state $\rf$ of the bath modes, 
\be
\tilde{\vro}(t=0)= \rd \otimes \rf. \label{factor}
\ee
Furthermore, we employ the Born-Markov approximation~\cite{breuer:os} and obtain
\begin{align}
\dot{\rd} & =  
-\mc{S}(\rd) + \text{h.c.}  , \label{master2}
\end{align}
where
\begin{align}
\mc{S}(\rd) & = \frac{1}{\hbar^2}\int\limits_0^{\infty} d\tau \text{Tr}_{\text{B}}\left\{
\left[V,e^{\mc{L}_{\text{B}} \tau}
V \rd \otimes \rf \right] \right\} .
\label{sint}
\end{align}
The application of the Born-Markov approximation requires that the coupling of 
the dark-state polaritons to bath excitations is sufficiently weak and in 
particular small as compared to the decay rate of bath excitations. 
Conditions for the validity of the Born-Markov approximation as well as the 
assumption in Eq.~(\ref{factor}) are discussed in Sec.~\ref{resmeq} and Appendix~\ref{bath}. 
In order to outline the evaluation of Eq.~(\ref{sint}), we write $V$ as 
\be 
V = V^{(+)} + V^{(-)}, 
\ee 
where the rising and lowering parts of $V$ are defined as 
\be 
V^{(+)} = \sum\limits_i B_i^{\dagger}S_i \,,\qquad
V^{(-)} = \sum\limits_i B_i S_i^{\dagger},
\ee
respectively. In this equation, $S_i$ represents a system operator comprised of 
dark-state polaritons, and $B_i$ is a bath operator. 
Since we assume that the bath is initially in its vacuum state, 
we have  $V^{(-)}\rf=0$ which allows us to replace 
$V\rf$ by $V^{(+)}\rf$ in Eq.~(\ref{sint}).  
In addition,  the second interaction Hamiltonian $V$ in 
Eq.~(\ref{sint}) can be replaced by  $V^{(-)}$ since the contribution of $V^{(+)}$ 
is negligible, see Appendix~\ref{bath}. We thus arrive at 
\begin{align}
\mc{S}(\rd) 
& = \frac{1}{\hbar^2}\int\limits_0^{\infty} d\tau \text{Tr}_{\text{B}}\left\{
\left[V^{(-)} , e^{\mc{L}_{\text{B}} \tau}
V^{(+)} \rd \otimes \rf \right] \right\}  \notag\\
& =  \frac{1}{\hbar^2}\sum\limits_{i,j}
\int\limits_0^{\infty} d\tau 
\text{Tr}_{\text{B}}\left\{B_i e^{\mc{L}_{\text{B}} \tau} B_j^{\dagger}\rf\right\}  \label{sint2} \\
& \hspace*{3cm}\times \left( S_i^{\dagger} S_j \rd - S_j \rd S_i^{\dagger}\right), \notag
\end{align}
and the evaluation of the bath correlation functions 
$\text{Tr}_{\text{B}}\{B_i \exp[\mc{L}_{\text{B}} \tau] B_j^{\dagger}\rf\}$ 
is  presented in Appendix~\ref{bath}. 
The final result for the  master equation~(\ref{master2}) is discussed in the next Section~\ref{resmeq}.
\section{MASTER EQUATION FOR DARK-STATE POLARITONS: RESULTS \label{resmeq}}
The master equation~(\ref{meq0}) in Sec.~\ref{description} 
describes the interaction of classical control and quantized probe fields 
with $N$ atoms. In Sec.~\ref{Smeq} we demonstrated that this master equation 
can be converted into a master equation for the reduced density 
operator $\rd$ of dark-state polaritons.  For equally strong control fields [see Eq.~(\ref{equalR})] 
and in the slow-light limit $\cos^2\theta\ll 1$ [see Eq.~(\ref{scO})], we obtain 
\begin{widetext}
\begin{align}
 \dot{\vro}_{\text{D}} = 
&  -\frac{i}{\hbar}[H_{\text{nd}},\rd] -\frac{i}{\hbar}[H_3 ,\rd] -\frac{i}{\hbar}[H_4 ,\rd]  
- \frac{\Gamma}{2\Omega_0^2}  \cos^2\theta \sum\limits_k \omega_k^2
\left( \psi_k^{\dagger} \psi_k \rd 
 +  \rd \psi_k^{\dagger} \psi_k - 
 2 \psi_k \rd  \psi_k^{\dagger}\right)    \label{meq4}\\
&  - \frac{\Gamma}{2\Omega_0^2} \Delta\omega^2  \cos^2\theta 
 \sum\limits_k \left( \psi_k^{\dagger}  \psi_k \rd + \rd  \psi_k^{\dagger}  \psi_k 
-2 \psi_k \rd \psi_k^{\dagger}\right)
\notag\\
& - \frac{g_2^2   \gamma_{42}/2}{ \Delta_{\theta}^2 + \gamma_{42}^2/4}\cos^2\theta \,  
\sum\limits_{k,p,q} 
\left( \psi_p^{\dagger} \psi_{k-p}^{\dagger}  \psi_q \psi_{k-q} \rd 
+ \rd \psi_p^{\dagger} \psi_{k-p}^{\dagger}  \psi_q  \psi_{k-q} 
- 2\psi_q \psi_{k-q} \rd \psi_p^{\dagger} \psi_{k-p}^{\dagger} \right) , \notag
\end{align}
\end{widetext}
where
\begin{align}
& H_{\text{nd}} = -\hbar\frac{\delta}{\Omega_0^2}\Delta\omega^2\, \cos^2\theta  
\sum\limits_k \psi_k^{\dagger}  \psi_k  , \label{hdb}
\end{align}
\begin{align}
&H_3 = -\hbar \frac{\delta}{\Omega_0^2}  \cos^2\theta \sum\limits_k \omega_k^2
 \psi_k^{\dagger}  \psi_k , \label{h3}\\
& H_4 = \frac{\hbar\Delta_{\theta}g_2^2}{ \Delta_{\theta}^2 + \gamma_{42}^2/4} \cos^2\theta 
\sum\limits_{k,p,q} 
 \psi_p^{\dagger} \psi_{k-p}^{\dagger}  \psi_q \psi_{k-q} . \label{h4}
\end{align}
Here $\Delta\omega=\omega_p-\omega_c$ 
is the frequency difference between the probe  and control fields,  
\begin{align}
\Delta_{\theta} = \Delta-\cot^2\theta\Delta\omega, \label{dtheta}
\end{align}
and $\Gamma=\gamma_{31}+\gamma_{32}$ is the full decay rate of state $\ket{3}$.  

Next we summarize the conditions under which Eq.~(\ref{meq4}) holds. 
First of all, we note that the two-photon detuning $\ve$ is constrained by Eq.~(\ref{twophot}), 
and it was  assumed that $\Omega_0$ defined in Eq.~(\ref{scO})  is large as compared 
to the decay rates of the excited states and the detuning with state $\ket{3}$,
\begin{align}
& \Omega_0 \gg \Gamma,\gamma_{42},|\delta|.  \label{approx1}
\end{align}
A key assumption in the derivation of Eq.~(\ref{meq4}) is that the number of atoms 
is much larger than the number of photons, and that initially almost all atoms 
are in state $\ket{1}$. This condition is a prerequisite for the bosonization 
described in Sec.~\ref{excitations}.   Furthermore, it was assumed in 
Sec.~\ref{elemination} that initially all bath modes are in the vacuum 
state, and that the initial density operator of the total system 
factorizes, see Eq.~(\ref{factor}).
These conditions can be met  if the initial dark-state polariton state is 
prepared via the slowing and stopping of a probe 
pulse~\cite{hau:99,kash:99,budker:99, liu:01,phillips:01,fleischhauer:00,fleischhauer:02,dey:03}. 
In this case, the initial state of the  dark-state polaritons is a slowly-varying spin coherence, 
and all other modes are in the vacuum state. The regime of stationary light can then be entered 
if the counter-propagating control fields are adiabatically switched on. 

The derivation of Eq.~(\ref{meq4}) relies on the validity of the 
Born-Markov approximation. 
The Born approximation employed in Sec.~\ref{elemination} requires that the coupling 
between the dark-state polaritons and other excitations is sufficiently weak and holds  
in the slow-light regime where $\cos^2\theta\ll 1$.  
On the other hand, 
the Markov approximation relies on the existence of two very different 
time scales. The fast time scale is represented by  the bath correlation times $T_{\text{B}}$ 
and is of the order of the lifetimes of the excited states $\ket{3}$ and $\ket{4}$. 
On the other hand, the slow time scale $T_{\text{S}}$ is given by the typical evolution time of 
dark-state polaritons. 
The condition $T_{\text{S}} \gg T_{\text{B}}$ which justifies the Markov approximation 
is fulfilled provided that the following inequalities hold,
\begin{align}
&\frac{\cos^2\theta c^2 k_{\text{max}}^2}{\Omega_0^2}\ll 1, 
&& \frac{2 |\delta|\cos^2\theta c^2 k_{\text{max}}^2}{\Omega_0^2\Gamma} \ll 1 , \label{markov1} \\
& \frac{16 g_2^2\cos^2\theta N_{\text{ph}}}{\gamma_{42}^2} \ll 1. \label{markov3}
\end{align}
Here $c$ is the speed of light in the fiber, 
$N_{\text{ph}}$ is the number of photons in the pulse and 
$k_{\text{max}}\ge|k|$ is the maximum of all occupied wave numbers. 
In addition, we emphasize that  the Markov approximation is only possible if we 
assume that the fast ground-state coherences $X_k(m)$ for $m\not=0$ [see Eq.~(\ref{xk})]  
are washed out  due to the atomic motion on a timescale comparable to 
the lifetime of the excited state $\ket{3}$, see Appendix~\ref{bath}. 
Note that this condition is not required in the case of the level scheme discussed in Sec.~\ref{other}.  
If $\Gamma_{\text{FO}}$ denotes the decay rate of the fast oscillating spin coherences, 
the validity of the master equation~(\ref{meq4}) requires 
\be
\frac{(\Omega_0 \cos\theta)^2}{\Gamma_{\text{FO}}}  \ll \Gamma.  \label{fcc}
\ee
If this condition is not fulfilled but if $\Gamma_{\text{FO}}$ is of the order of 
the decay rates of the excited states, then the general structure of Eq.~(\ref{meq4}) 
remains the same, but the pre-factors of 
$H_3$, $H_{\text{nd}}$ and the decay terms proportional to $\Gamma$ will be different.

Finally, we note that the Hamiltonian $H_{\text{nd}}$ in Eq.~(\ref{hdb}) 
represents a constant energy shift of the polariton excitations. This term 
is only present if $\Delta\omega\not=0$ and hence if the ground 
states $\ket{1}$ and $\ket{2}$ are non-degenerate. In the following, 
we will  work in an interaction picture  with respect to $H_{\text{nd}}$, 
and the representation of the master equation in this rotating frame can 
be obtained if $H_{\text{nd}}$ is omitted in Eq.~(\ref{meq4}). 
\subsection{STATIONARY LIGHT \label{statlight}}
In this Section we focus on the phenomenon of stationary light~\cite{andre:02,bajcsy:03} 
that arises from the interaction of the probe and
control fields  with the $\Lambda$ system formed by states $\ket{1}$, 
$\ket{2}$ and $\ket{3}$. Formally, the reduction of the master equation~(\ref{meq4}) 
to this case is accomplished if the coupling constant $g_2$ is set equal to zero. 
In the following, we   formulate the master equation in terms 
of the  operator 
\be
\psi(z) = \frac{1}{\sqrt{L}}\sum_k e^{i k z} \psi_k . \label{fop}
\ee
It follows from   Eq.~(\ref{bcd}) that 
$\psi$ is a bosonic field operator that obeys the commutation relations
\be
[\psi(z),\psi^{\dagger}(z^{\prime})]=\delta(z-z^{\prime}),\quad 
[\psi(z),\psi(z^{\prime})]=0.
\ee
The master equation~(\ref{meq4})  for $g_2=0$ reads 
\begin{align}
 \dot{\rd} = &   -\frac{i}{\hbar}[H_3 ,\rd] + \mc{L}_1\rd + \mc{L}_2\rd,  \label{meqSL}
\end{align}
where $H_3$ is defined in Eq.~(\ref{h3}) and shows 
that the polaritons experience a quadratic dispersion relation. With the 
definition~(\ref{fop}), this Hamiltonian can be written in the form of  a kinetic 
energy term, 
\begin{align}
& H_3 =  \frac{\hbar^2}{2 m_{\text{eff}}} \int\limits_0^L \text{d}z
 \partial_z\psi^{\dagger}  \partial_z\psi   , \label{h3f}
\end{align}
where 
\begin{align}
& m_{\text{eff}}=-\frac{\hbar\Omega_0^2}{2\delta c^2 \cos^2\theta} 
\end{align}
is the effective mass of the polaritons. 

The terms $\mc{L}_1\rd$ and $\mc{L}_2\rd$ in Eq.~(\ref{meqSL}) describe polariton losses 
and are defined as 
\begin{align}
&\mc{L}_1 \rd =  -\frac{\Gamma}{2\Omega_0^2} c^2 \cos^2\theta  \mc{D}[\partial_z\psi], \label{l1}\\
& \mc{L}_2 \rd = -\frac{\Gamma}{2\Omega_0^2} \Delta\omega^2  \cos^2\theta \mc{D}[\psi], \label{l2}
\end{align}
respectively, where 
\begin{align}
\mc{D}[\hat{X}]=\int\limits_0^L \text{d}z
(\hat{X}^{\dagger}\hat{X}\rd+\rd\hat{X}^{\dagger} \hat{X}-2\hat{X}\rd\hat{X}^{\dagger})
\end{align}
is a dissipator in Lindblad form~\cite{breuer:os} for an operator $\hat{X}$.  
\begin{figure}[t!]
\bc
\includegraphics[scale=1]{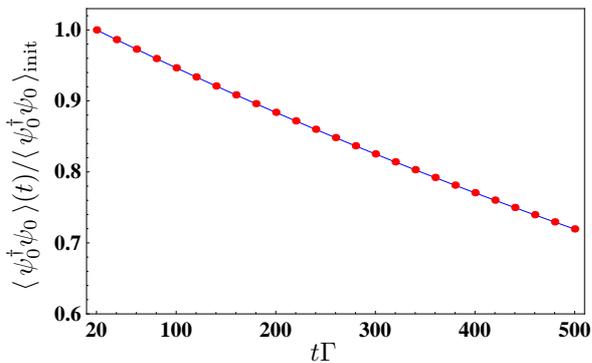}
\ec
\caption{\label{fig3} 
(Color online) Decay of the $k=0$ polariton mode for non-degenerate ground states $\ket{1}$ and $\ket{2}$. 
The solid line shows the decay according to Eq.~(\ref{dzero}), the corresponding result 
from a numerical integration of Maxwell-Bloch equations is indicated by red dots. 
The parameters are $\Delta\omega=150 \Gamma$, $\cos^2\theta= 2.5\times 10^{-4}$, 
$\Omega_0= 90\Gamma$,  $\delta=0$ and $\ve$  is fixed by Eq.~(\ref{twophot}). 
The initial spin coherence $R_{21}$ (see Appendix~\ref{mbe})  has a Gaussian shape 
$R_{21} \propto \exp[-(z-z_0)^2/(2\sigma^2)]$ with $\sigma=24.3 \Gamma c/\Omega_0^2$, 
and the control fields are ramped up according to 
$\Omega_c(t) =  0.5[1+\tanh(t-10/\Gamma)]\times \Gamma$. The initial value 
$\mean{\psi_0^{\dagger}\psi_0}_{\text{init}}$ refers to $\mean{\psi_0^{\dagger}\psi_0}$ at 
$t=20/\Gamma$ where the control fields  have attained their maximal value of $\Omega_c=\Gamma$. 
}
\end{figure}
%
The term $\mc{L}_1\rd$ arises due to the   coupling between  dark-state polaritons 
and the difference mode $D_k$. Since the decay rate of the individual modes 
increases quadratically with the wave number, $\mc{L}_1\rd$ does not 
result in an exponential damping but leads to diffusion~\cite{bajcsy:03,zimmer:06}.  
On the contrary, $\mc{L}_2\rd$ leads to identical  decay rates for all modes. 
This term stems from the coupling  between dark-state polaritons $\psi_k$ and  bright polaritons $\phi_k$. 
Since this loss mechanism has not been discussed in the literature yet, 
we investigate it in more detail here. 
First of all, we note that $\mc{L}_2\rd$ is proportional to  $\Delta\omega^2$, 
and $\Delta\omega$ 
practically coincides with the splitting of the ground states 
$\ket{1}$ and $\ket{2}$ [see Eq.~(\ref{do})]. 
An important consequence of $\mc{L}_2\rd$ is that in contrast to EIT,  
dark-state polaritons in the $k=0$ mode  decay under the conditions of 
stationary light provided that the ground states $\ket{1}$ and $\ket{2}$ are non-degenerate. 
The decay of the mean number of dark-state 
polaritons in the $k=0$ mode can be calculated from Eq.~(\ref{meqSL}) and is given by 
\begin{align}
 \partial_t \mean{\psi_0^{\dagger}\psi_0}  =   
- \frac{\Gamma}{\Omega_0^2} \Delta\omega^2  \cos^2\theta \mean{\psi_0^{\dagger}\psi_0} . \label{dzero}
\end{align}
The accuracy of this  result can be confirmed numerically if $\mean{\psi_0^{\dagger}\psi_0}$ 
is evaluated via  Maxwell-Bloch equations for classical probe fields (see Appendix~\ref{mbe}).  
The result is shown in Fig.~\ref{fig3}, where the solid line corresponds to 
the exponential decay according to Eq.~(\ref{dzero}). The dotted line represents 
$\mean{\psi_0^{\dagger}\psi_0}$ obtained from the numerical integration of Maxwell-Bloch equations 
and is in perfect agreement with the predictions of the master equation~(\ref{meqSL}). 
%

\begin{figure}[t!]
\bc
\includegraphics[scale=1]{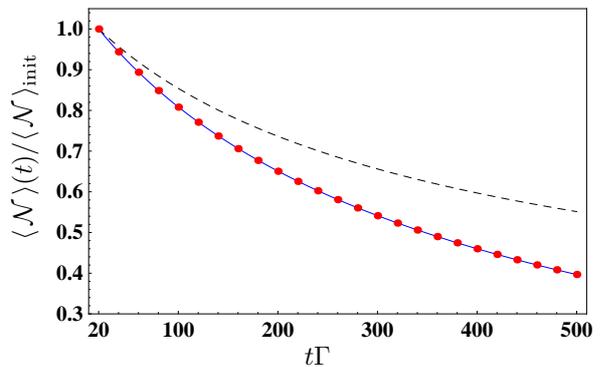}
\ec
\caption{\label{fig4} 
(Color online) The number of dark-state polaritons is shown as a function of time.
The results of  Eqs.~(\ref{tloss}) and Eq.~(\ref{em}) are  represented by the 
solid and the dotted line, respectively.  
The good agreement indicates that the polariton losses are described correctly by 
the master equation~(\ref{meqSL}). 
The number of polaritons follows the dashed 
line if  the additional decay term 
$\mc{L}_2\rd$ in Eq.~(\ref{meqSL}) is omitted. 
The parameters are the same as in Fig.~\ref{fig3}. 
}
\end{figure}
%
Next we compare the impact of the loss terms $\mc{L}_1\rd$ and $\mc{L}_2\rd$. 
Equations~(\ref{l1}) and~(\ref{l2}) imply  that $\mc{L}_2\rd$ becomes 
comparable to $\mc{L}_1\rd$
if $|\Delta\omega|$ is of the order of $c k_{\text{max}}$. Since 
$c k_{\text{max}}\approx |\Delta\omega| + c \sigma_{k}$ where $\sigma_{k}$ 
is the width of the polariton pulse in $k$ space, the two loss terms 
are comparable if 
\be
|\Delta\omega|\ge c\sigma_{k}.\label{ineq}
\ee 
 On the other hand, 
the width  $\sigma_{k}$ can be estimated to be of the order of $2\pi /L$, 
where $L$ is the length of the system.  The  inequality~(\ref{ineq}) thus 
implies that the impact of  
$\mc{L}_2\rd$ is of the same order of  $\mc{L}_1\rd$ if the wavelength associated with 
the beat note $\Delta\omega$ of the probe and control fields is comparable or 
shorter than $L$.
For realistic values of  $L$ of a few centimeters, the term $\mc{L}_2\rd$ 
will have a significant impact  if  $|\Delta\omega|$  is of the order of a few GHz 
or larger. 
Note that polariton losses can be minimized by minimizing $|\Delta\omega|$. 
This is obvious for $\mc{L}_2\rd$ since it is proportional to $\Delta\omega^2$. 
However, also the impact of   $\mc{L}_1\rd$ increases with 
increasing values of $|\Delta\omega|$ since 
$c k_{\text{max}}\approx |\Delta\omega| + c \sigma_{k}$. 

The total losses of dark-state polaritons can be calculated from  Eq.~(\ref{meqSL}). 
We find that the mean number of dark-state polaritons $\mean{\mc{N}}$ obeys
\begin{align}
 \partial_t \mean{\mc{N}} = -\frac{\Gamma}{\Omega_0^2}\cos^2\theta
\left(\Delta\omega^2  
+  \frac{1}{\mean{\mc{N}}} \sum\limits_k \omega_k^2 \mean{\psi_k^{\dagger}\psi_k} \right) \mean{\mc{N}}, 
\label{tloss}
\end{align}
where $\mc{N} = \sum_k \psi_k^{\dagger}\psi_k$ is the polariton number operator. 
The solid line in Fig.~\ref{fig4} shows the losses of polaritons according to Eq.~(\ref{tloss})   
where $\mean{\psi_k^{\dagger}\psi_k}$ was calculated via the numerical integration 
of Maxwell-Bloch equations, see Appendix~\ref{mbe}.   
On the other hand, the number of dark-state polaritons is proportional to 
the electromagnetic field intensity, 
\begin{align}
\mean{\mc{N}}\propto \int\limits_0^L \text{d}z \left(|\mc{G}_+|^2 + |\mc{G}_-|^2 \right) , 
\label{em}
\end{align}
where $\mc{G}_+$  ($\mc{G}_-$) is the Rabi frequency of the classical 
probe field propagating in the positive (negative) $z$ direction.  
In order to test Eq.~(\ref{tloss}), we evaluate the  right-hand side of Eq.~(\ref{em}) 
as a function of time 
from a numerical integration of  Maxwell-Bloch equations. 
The result is shown as the dotted line in Fig.~\ref{fig4}  
and in excellent agreement with the findings of  Eq.~(\ref{tloss}). 
Finally, the dashed line in Fig.~\ref{fig4} represents the polariton losses if the 
term $\mc{L}_2\rd$ were neglected and shows that $\mc{L}_2\rd$ contributes significantly. 
Note that the parameters of the  first stationary light experiment~\cite{bajcsy:03} 
indicate a  ratio of  $|\Delta\omega|/(c k_{\text{max}}) \approx 0.75$, which is even larger 
than the value of $|\Delta\omega/(c k_{\text{max}}) \approx 0.25$ chosen in  Fig.~\ref{fig4}. 

We emphasize that the loss term $\mc{L}_2\rd$  arises only in the presence of 
two counter-propagating control fields. In this case, the photonic component 
$A_k$ in Eq.~(\ref{ak}) of the dark-state polaritons $\psi_k$ 
is comprised of counter-propagating probe field modes that are grouped around 
the wave numbers $\pm k_c$ of the control field rather than the probe field, see Sec.~\ref{DSP}. 
It follows that the total Hamiltonian 
$H = H_0 + H_{\Lambda} + H_{\text{NL}}$ in Sec.~\ref{description}  does not possess a 
true dark state for $\Delta\omega\not=0$. Even the dark-state polaritons $\psi_k$ 
in the $k=0$ mode experience a coupling to bright-state polaritons $\phi_k$ and  thus decay. 
This  mechanism is at the heart of Eq.~(\ref{dzero}) that describes the loss of 
dark-state polaritons in the $k=0$ mode. 
The situation is different in a standard EIT configuration where only one pair of co-propagating 
probe and control fields is present, 
and the loss of dark-state polaritons is described by $\mc{L}_1\rd$ only.

\subsection{STATIONARY LIGHT WITH TWO-PARTICLE INTERACTION \label{nonlin}}
Next we restate the full master equation~(\ref{meq4})  in terms of the field 
operators $\psi(z)$ defined in Eq.~(\ref{fop}). In addition to the  terms 
discussed in Sec.~\ref{statlight}, we have to take into account all contributions 
proportional to the coupling constant $g_2$ in Eq.~(\ref{meq4}) that 
account for elastic and inelastic polariton-polariton interactions. 
We obtain~\cite{kiffner:10}
\be 
\hbar \dot{\rd} = -i H_{\text{eff}}\rd +i \rd H_{\text{eff}}^{\dagger} + \mc{I}\rd 
+ \hbar \mc{L}_1\rd + \hbar \mc{L}_2\rd, 
\label{meq}
\ee
where $H_{\text{eff}}$ is a non-hermitian Hamiltonian, 
\be
H_{\text{eff}} = H_3 
+  \frac{\tilde{g}}{2}\int_0^L \text{d}z  \psi^{\dagger 2}  \psi^2\,,
\label{Heff}
\ee
and $H_3$, $\mc{L}_1\rd$ and $\mc{L}_2\rd$ are defined in Eqs.~(\ref{h3f}),~(\ref{l1}) and~(\ref{l2}), 
respectively. The parameter 
\begin{align}
\tilde{g}=\frac{2\hbar L g_2^2\cos^2\theta}{\Delta-\cos^2\theta \Delta\omega + i \gamma_{42}/2}
\label{coupling}
\end{align}
is the complex coupling constant, and 
\begin{align}
& \mc{I}\rd = - \text{Im}(\tilde{g}) \int\limits_0^L \text{d}z \psi^2 \rd  \psi^{\dagger 2} .
\label{ird}
\end{align}
The  term  proportional to $\tilde{g}$ in Eq.~(\ref{Heff}) and 
$\mc{I}\rd$ in Eq.~(\ref{ird}) account for elastic and inelastic two-particle interactions that 
originate from   the coupling of dark-state polaritons to the 
excited state $\ket{4}$. More precisely, the real part of $\tilde{g}$ gives 
rise to a hermitian contribution to $H_{\text{eff}}$ that accounts for 
elastic two-particle collisions. On the other hand, the imaginary part of $\tilde{g}$ 
together with $\mc{I}\rd$ gives rise to a two-particle loss term that can be written 
in Lindblad form as  $\text{Im}(\tilde{g}/2)\mc{D}[\psi^2]$. 

The master equation~(\ref{meq}) is  equivalent to Eq.~(\ref{meq4}) and 
describes a one-dimensional system of bosons with effective mass $m_{\text{eff}}$ that 
experience elastic and inelastic two-particle interactions. Except for 
the two loss terms $\mc{L}_1\rd$ and $\mc{L}_2\rd$, Eq.~(\ref{meq}) can 
be identified with the dissipative Lieb-Liniger model discussed in the next Section.

\subsection{DISSIPATIVE LIEB-LINIGER MODEL \label{dll}}
The original Lieb-Liniger model~\cite{lieb:63} established in 
1963 describes bosons in one dimension 
that experience a repulsive contact interaction. In the limit of strong interactions, 
the bosons can enter the regime of a Tonks-Girardeau gas~\cite{girardeau:60} where 
they behave with respect to many observables as if they were fermions. 
Recently, it was shown~\cite{duerr:09} that the original Lieb-Liniger model 
can be generalized to systems where the bosons experience a 
contact interaction with complex coupling constant, i.e.,  
they undergo elastic or inelastic two-particle interactions. 
This dissipative Lieb-Liniger model~\cite{duerr:09} shows that 
even a purely dissipative interaction effectively results in 
a repulsion and produces a Tonks-Girardeau gas  in the limit of strong interactions.

The master equation~(\ref{meq}) 
can be identified with the dissipative Lieb-Liniger model provided that the loss 
terms $\mc{L}_1\rd$ and $\mc{L}_2\rd$ are negligible. In the following we 
specify the conditions that justify this approximation 
and  assume that 
$\Delta\omega^2$ is small enough such that the impact of $\mc{L}_2\rd$ is small as compared to 
$\mc{L}_1\rd$, see Section~\ref{statlight}. 
On the other hand, the diffusion term $\mc{L}_1\rd$   is negligible if 
two conditions are met. First, the dynamics induced by the kinetic energy term 
proportional to $m_{\text{eff}}$ in Eq.~(\ref{Heff}) must 
be fast as compared to the inverse decay rate of polaritons introduced by $\mc{L}_1\rd$. This can  
be achieved if we set $|\delta|\gg \Gamma$. 
Second, losses due to $\mc{L}_1\rd$ must be negligible which imposes 
a limit on the maximal evolution time 
\begin{align}
t_{\text{max}} \ll  \frac{1}{\Gamma}\frac{1}{\cos^2\theta}\frac{2\Omega_0^2}{ c^2 k_{\text{max}}^2} . 
\label{tlimit}
\end{align} 
Note  that $t_{\text{max}}\gg 1/\Gamma$ is much larger than the lifetime of the excited 
state $\ket{3}$. 
Under these conditions, the master equation~(\ref{meq}) reduces to 
\be
\hbar \dot{\rd} = -i H_{\text{eff}}\rd +i \rd H_{\text{eff}}^{\dagger} + \mc{I}\rd
\ee
and  can be identified with  
the generalized Lieb-Liniger model~\cite{duerr:09} for a  one-dimensional 
system of bosons with  mass $m_{\text{eff}}$ and complex interaction parameter $\tilde{g}$. 
All features of the Lieb-Liniger model~\cite{lieb:63,duerr:09}   
are characterized by a single, dimensionless parameter 
\be
G = \frac{m_{\text{eff}} \tilde{g}}{\hbar^2 N_{\text{ph}}/L},
\ee
where $N_{\text{ph}}$ 
is the number of photons in the pulse. 
In the strongly correlated regime $|G|\gg 1$, the interaction between the particles creates 
a Tonks-Girardeau (TG) gas where polaritons behave like impenetrable hard-core particles that 
never occupy the same position. 
Depending on the sign of the detuning $\delta$ and $\Delta$, the elastic interaction 
between the polaritons can be either attractive or repulsive. The dissipative component 
of the interaction is negligible for $\Delta\gg\gamma_{42}$. In this case, 
the preparation of a TG gas of polaritons with repulsion can be achieved if 
$\delta \Delta<0$~\cite{chang:08}. Note that the interaction 
becomes attractive if $\delta \Delta > 0$ which opens up the 
possibility to enter the super Tonks-Girardeau regime for 
polaritons~\cite{astrakharchik:05,batchelor:05,haller:09}. 
Since  the coupling constant $\tilde{g}$ in Eq.~(\ref{coupling}) is 
maximal for  $\Delta=0$, 
the Lieb-Liniger parameter $|G|$ and hence the induced correlations are maximal 
for purely dissipative interactions~\cite{kiffner:10}. 

\subsection{OTHER REALIZATIONS \label{other}}
\begin{figure}[t!]
\includegraphics[scale=1]{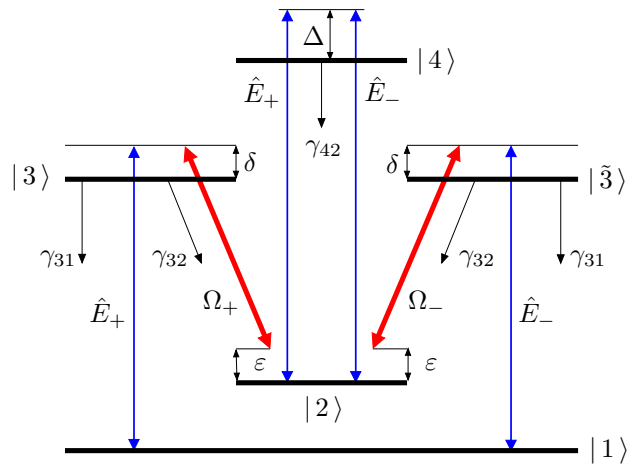}
\caption{\label{fig5} 
(Color online) Alternative level scheme that gives rise to the 
master equation for polaritons described in Sec.~\ref{resmeq}. 
All symbols are  defined in the same way as in Fig.~\ref{fig2}. 
}
\end{figure}

The master equation for dark-state polaritons in Sec.~\ref{resmeq} was derived 
under the assumption that the level scheme of each atom is given by Fig.~\ref{fig2}. 
Here we point out that stationary light and interacting dark-state polaritons can 
be realized as well with  the level scheme in Fig.~\ref{fig5} that was  
suggested in~\cite{zimmer:08,fleischhauer:08}. 
Moreover, a  straightforward modification of the formalism described in Sec.~\ref{Smeq} 
demonstrates that the level scheme of Fig.~\ref{fig5} leads to 
the same master equation~(\ref{meq4})  for dark-state polaritons as 
the configuration in  Fig.~\ref{fig2}. 
On the contrary, each system displays characteristic  advantages and disadvantages 
that we discuss now. 

%

The major difference between the two configurations  
is that  the level scheme in Fig.~\ref{fig5} 
creates stationary light via the double-$\Lambda$ system 
formed by states $\ket{1}$, $\ket{2}$, $\ket{3}$ and $\ket{\tilde{3}}$. 
Here each probe field interacts  only with the co-propagating  control field,  
and thus no fast oscillating spin coherences $X_k(m)$ for $m\not=0$ are 
produced~\cite{comment2}. 
This feature is a  significant advantage of the system shown in Fig.~\ref{fig5}, since  
it implies that  the master equation~(\ref{meq4}) remains valid for ultracold atoms or  
stationary atoms where the condition~(\ref{fcc}) cannot be fulfilled. 

On the other  hand, the implementation of the configuration in 
Fig.~\ref{fig5} comes along with difficulties that do not occur 
in the case of the level scheme in Fig.~\ref{fig2}.  
First, we note that the transitions in the $\Lambda$ configuration of Fig.~\ref{fig2} 
can be selected by  polarization, even if the ground states $\ket{1}$ and 
$\ket{2}$ are degenerate. This is not the case for the level scheme 
in Fig.~\ref{fig5} where 
the level splitting between the ground states 
$\ket{1}$ and $\ket{2}$ must be large enough such that 
the transitions $\ket{2}\leftrightarrow\ket{3}$, $\ket{1}\leftrightarrow\ket{3}$ 
and $\ket{2}\leftrightarrow\ket{\tilde{3}}$, $\ket{1}\leftrightarrow\ket{\tilde{3}}$ 
can  be addressed independently. Consequently, the parameter 
$|\Delta\omega|$ must be significantly larger than the Rabi frequencies 
$\Omega_{\pm}$ of the control fields which leads to additional 
losses, see Sec.~\ref{resmeq}. Second, we point out that the initial 
polariton pulse in $k$ space is centered around $\Delta\omega/c$ 
if it is prepared via the slowing and stopping of a probe pulse. 
The reason for this is that the wave numbers $k$ of stationary-light polaritons 
have to be grouped around the wavenumbers $\pm k_c$ of the control fields, see 
Sec.~\ref{DSP}. 
If the kinetic energy of the stationary-light polaritons is different 
from zero (i.e., $\delta\not=0$) and in the case of the level 
scheme of Fig.~\ref{fig5},  the mandatory choice of 
$\Delta\omega\not=0$ will lead to a moving polariton pulse even 
if the control fields have the same intensity.

\section{SUMMARY \label{summary}}
In this paper we   introduced a  technique for the description of 
light-matter interactions under conditions of EIT. More specifically, we 
described a general method for the derivation of a master equation for dark-state 
polaritons. 
In contrast to the standard description~\cite{fleischhauer:00,fleischhauer:02} 
based on a Heisenberg-Langevin approach, our master equation facilitates the treatment 
of polariton losses.  This achievement allows us  to model 
general polariton-polariton interactions that may be conservative, dissipative 
or a mixture of both.
In particular, the master equation approach enables us to study dissipation-induced  
correlations~\cite{syassen:08,duerr:09} that are promising  in the quest for highly correlated 
systems since they can be considerably stronger than their conservative counterparts~\cite{kiffner:10}. 
For the illustration of our technique we use the example of stationary light 
polaritons that experience a conservative or dissipative interaction. 
The resulting master equation is discussed in various limiting cases. 
For stationary-light polaritons we find an additional loss mechanism
that was overlooked so far. It is related to the fact 
that the total Hamiltonian of the system does not possess a true dark state 
if the ground states are non-degenerate. 
In particular,  polariton losses in level configurations with non-degenerate 
ground states can be a multiple of those in level schemes 
with degenerate ground states. 
Furthermore, we specified conditions that allow us to reduce the full 
master equation for dark-state polaritons to the dissipative Lieb-Liniger model~\cite{duerr:09}. 
Finally, we discussed the atomic level scheme in Fig.~\ref{fig5} that 
leads to the same master equation for dark-state polaritons as the one in Fig.~\ref{fig2} and 
compared the advantages and disadvantages of both configurations. 

At the heart of our approach is a mapping of the full system dynamics 
to a conceptually simple system of coupled bosonic modes.
This mapping could be the starting point for future studies of EIT systems 
beyond a Markovian master equation for dark-state polaritons.

\begin{acknowledgments}
This work is part of the Emmy Noether project 
HA 5593/1-1 funded by the German Research Foundation (DFG).
\end{acknowledgments}

\appendix
\section{REPRESENTATION OF OPERATORS IN $\hf$ \label{representation}}
Here we show how the master equation Eq.~(\ref{meq0}) can be expressed in terms of 
bosonic creation and annihilation operators if the system dynamics is restricted 
to the subspace $\hf$. To this end, 
let $\hat{X}$  be an operator acting on the total state space $\hc$. 
In the following, we describe a procedure that allows one to construct 
an operator $\hat{X}_{O}$ that comprises 
of  bosonic operators $O_i\in\mc{O}$ [see Eq.~({\ref{operators})]  
and that coincides approximately with $\hat{X}$ in the subspace $\hf$. 
The latter condition implies that $\hat{X}_{O}$ and $\hat{X}$  
must necessarily obey the same commutation relations with 
the creation operators $O_i^{\dagger}$   ($O_i\in\mc{O}$) in $\hf$, 
\begin{align}
& \left[\hat{X}, O_i^{\dagger}\right]_{\hf} 
= \left[\hat{X}_{O}, O_i^{\dagger}\right]_{\hf}. \label{cond}
\end{align}
Furthermore, we have $\hat{X}_{O}\ket{s}=\hat{X}\ket{s}$ 
for an arbitrary state $\ket{s}\in\hf$. 
In all situations considered below, the operator $\hat{X}$ annihilates 
the vacuum state,  $\hat{X}\ket{0} = 0$. In addition to 
Eq.~(\ref{cond}), we thus require 
\be
\hat{X}_{O}\ket{0} = 0\,.
\label{assump}
\ee
The two conditions in Eqs.~(\ref{cond}) and~(\ref{assump}) 
are sufficient to determine $\hat{X}_{O}$ since they guarantee 
that the matrix elements of 
$\hat{X}$ and $\hat{X}_{O}$  are identical in the subspace $\hf$. 
In order to see this, we evaluate the action of $\hat{X}$ applied 
to the  states $\ket{\{n_1,\ldots,n_M\}}$ that span $\hf$, 
\be 
\hat{X} \ket{\{n_1,\ldots,n_M\}} 
= \left[\hat{X},\prod_{i=1}^{M}\frac{1}{\sqrt{n_i!}}\big(\hat{O}_i^{\dagger}\big)^{n_i}\right]\ket{0}\,.
\label{comm}
\ee
Here we employed the definition~(\ref{states}) and the relation $\hat{X}\ket{0} = 0$. 
The recursive application of the identity 
$[A,B C] = [A,B] C + B [A , C]$
allows us to write the commutator on the right-hand side of Eq.~(\ref{comm}) as a sum of 
terms where only the commutator of $\hat{X}$ with one of the  creation operators  
$O_i^{\dagger}$ appears. 
Therefore,  Eqs.~(\ref{cond}) and~(\ref{assump}) guarantee that the matrix elements of 
$\hat{X}$ and $\hat{X}_{O}$  are identical in the subspace $\hf$. 

As an example, we discuss the representation of $\sum_{\mu=1}^N A_{22}^{(\mu)}$ in terms 
of the operators $O_i \in\mc{O}$. The only non-vanishing commutators are 
\be 
\left[\sum\limits_{\mu=1}^N   A_{22}^{(\mu)}, X_k^{\dagger}(m)\right] = X_k^{\dagger}(m), 
\label{a22}
\ee 
and we have $\sum_{\mu=1}^N A_{22}^{(\mu)}\ket{0}=0$. According to 
Eqs.~(\ref{cond}),~(\ref{assump}) 
and with the bosonic commutation relations obeyed by the operators $X_k(m)$, 
the representation of  $\sum_{\mu=1}^N A_{22}^{(\mu)}$ in  $\hf$ is given by 
\begin{align}
 \sum\limits_{\mu=1}^N   A_{22}^{(\mu)} =
\sum\limits_{m}\sum\limits_k  X_k^{\dagger}(m) X_k(m) . \label{a22new}
\end{align}
The representation of the remaining operators that appear in 
the Hamiltonian $H$ in Eq.~(\ref{meq0}) can 
be found in a similar way, the result is 
\begin{align}
& \sum\limits_{\mu=1}^N   A_{33}^{(\mu)} =
\sum\limits_{m}\sum\limits_k  H_k^{\dagger}(m) H_k(m) , \label{a33}  \\
&   \sum\limits_{\mu=1}^N   A_{44}^{(\mu)} =
\sum\limits_{m}\sum\limits_k  I_k^{\dagger}(m) I_k(m) ,  \label{a44}  \\
& \sum\limits_{\mu=1}^N \Si{32}{\mu} e^{i k_c z_{\mu}} = 
\sum\limits_m \sum\limits_k X_k(m) H_k^{\dagger}(m+1), \\
& \sum\limits_{\mu=1}^N \Si{32}{\mu} e^{-i k_c z_{\mu}} = 
\sum\limits_m \sum\limits_k X_k(m) H_k^{\dagger}(m-1) ,  
\end{align}
\begin{align}
& \sum\limits_{\mu=1}^N \Si{42}{\mu} e^{i (k_c+k) z_{\mu}} = 
\sum\limits_m \sum\limits_p X_{p-k}(m) I_p^{\dagger}(m+1) , \\
& \sum\limits_{\mu=1}^N \Si{42}{\mu} e^{-i(k_c-k) z_{\mu}} = 
\sum\limits_m \sum\limits_p X_{p-k}(m) I_p^{\dagger}(m-1) .  
\end{align}
These relations together with the inverse relations of Eqs.~(\ref{ak}) and~(\ref{dk})   
allow us to find the representation of $H_0$, $H_{\Lambda}$ and $H_{\text{NL}}$ in $\hf$, 
\begin{widetext}
\begin{align}
  \tilde{H}_0 =  
& -\hbar \Delta\omega \sum_{k}\left(A_k^{\dagger}A_k+D_k^{\dagger}D_k\right) 
 + 2 \hbar  \sin\varphi\cos\varphi 
\sum_{k}\omega_k \left(A_k^{\dagger}D_k + A_k D_k^{\dagger}\right) \label{h0fe}\\
& -\hbar \varepsilon \sum\limits_{m}\sum\limits_k  X_k^{\dagger}(m) X_k(m) 
 -\hbar \delta \sum\limits_{m}\sum\limits_k  H_k^{\dagger}(m) H_k(m) 
 -\hbar (\Delta + \varepsilon) \sum\limits_{m}\sum\limits_k  I_k^{\dagger}(m) I_k(m) \\
& -\hbar (\sin^2\varphi-\cos^2\varphi)\sum_{k}\omega_k \left(D_k^{\dagger}D_k - A_k^{\dagger}A_k \right),\notag 
\end{align}
\begin{align}
\tilde{H}_{\Lambda}   =   &
- \hbar\Omega_0\sin\theta\sum\limits_k \left\{
A_k \left[H_k^{\dagger}(1)\sin\varphi +H_k^{\dagger}(-1)\cos\varphi \right] 
 + D_k\left[H_k^{\dagger}(1)\cos\varphi - H_k^{\dagger}(-1) \sin\varphi \right]\right\}  \label{hlambdafe}\\
&- \hbar\Omega_0\cos\theta\sum\limits_m \sum\limits_k X_k(m) \left[ 
 H_k^{\dagger}(m+1) \sin\varphi
 +  H_k^{\dagger}(m-1) \cos\varphi \right] +\text{h.c.} , \notag 
\end{align}
\begin{align}
 \tilde{H}_{\text{NL}} = &  
- \hbar g_2 \sum\limits_k A_k \sum\limits_m \sum\limits_p X_{p-k}(m) 
\left[  I_p^{\dagger}(m+1)  \sin\varphi
 +   I_p^{\dagger}(m-1) \cos\varphi \right]  \label{hnlfe}\\
& - \hbar g_2 \sum\limits_k D_k \sum\limits_m \sum\limits_p  X_{p-k}(m)
\left[ I_p^{\dagger}(m+1) \cos\varphi
 -   I_p^{\dagger}(m-1) \sin\varphi \right] + \text{h.c.} . \notag 
\end{align}
For the full representation of the master equation in $\hf$ it remains to 
transform the decay term $\mc{L}_{\gamma}\rd$, see Eq.~(\ref{decay0}). 
The representation of the terms that describe the decay of the excited states in Eq.~(\ref{decay0}) can 
be found via Eqs.~(\ref{a33}) and~(\ref{a44}). 
Since the super-operator $\tilde{\mc{L}}_{\gamma}\tilde{\vro}$ 
must preserve the trace of the density operator, we find 
\begin{align}
\tilde{\mc{L}}_{\gamma}\tilde{\vro} = 
& - \frac{\Gamma}{2} \sum\limits_{m}\sum\limits_k
\left[ H_k^{\dagger}(m)H_k(m) \tilde{\vro}  
 + \tilde{\vro} H_k^{\dagger}(m)H_k(m) -2 H_k(m) \tilde{\vro} H_k^{\dagger}(m) \right] \label{decay1} \\
& -\frac{\gamma_{42}}{2} \sum\limits_{m}\sum\limits_k
\left[ I_k^{\dagger}(m)I_k(m) \tilde{\vro}  
 + \tilde{\vro} I_k^{\dagger}(m)I_k(m)   -2 I_k(m) \tilde{\vro} I_k^{\dagger}(m) \right] ,  
\notag
\end{align}
where $\Gamma = \gamma_{31} + \gamma_{32}$ is the full decay rate of the excited state $\ket{3}$. 
\section{DEFINITIONS \label{definitions}}
The master equation~(\ref{meq2}) is obtained from~(\ref{meq1}) for $\Omega_+=\Omega_-$ and 
if we exchange the set of operators $\{A_k,X_k(0),H_k(1),H_k(-1),I_k(1),I_k(-1)\}$ 
by the new set $\{\psi_k,\phi_k,P_k,Q_k,U_k,V_k\}$ according to 
\begin{align}
& \psi_k = A_k \cos\theta - X_k(0) \sin\theta, &&
\phi_k = A_k \sin\theta + X_k(0) \cos\theta , \label{r1} \\
& P_k = [H_k(1)+H_k(-1)]/\sqrt{2},  &&
Q_k = [H_k(1)-H_k(-1)]/\sqrt{2}, \\
& U_k = [I_k(1)+I_k(-1)]/\sqrt{2}, &&
 V_k = [I_k(1)-I_k(-1)]/\sqrt{2}. \label{r3}
\end{align}
Since the old and new operators are related by a unitary transformation, the 
new operators $\{\psi_k,\phi_k,P_k,Q_k,U_k,V_k\}$ obey bosonic commutation relations. 
If the inverse  relations of Eqs.~(\ref{r1})-(\ref{r3}) are plugged into Eqs.~(\ref{h0fe}),~(\ref{hlambdafe}) 
~(\ref{hnlfe}) and~(\ref{decay1}), we obtain the master equation~(\ref{meq2}).  Furthermore, we employ the 
condition in Eq.~(\ref{twophot})  and the definition $\Delta_{\theta} = \Delta-\cot^2\theta\Delta\omega$. 
The bath Hamiltonian in Eq.~(\ref{lr}) is comprised of four parts,
\be
H_{\text{B}} = H_{\text{B}}^{(0)} + H_{\text{B}}^{(1)}+ H_{\text{B}}^{(2)} + H_{\text{B}}^{(3)}, 
\qquad \text{where}
\ee
\begin{align}
H_{\text{B}}^{(0)} = & -\hbar \Delta\omega \sum\limits_{k} \left(
 D_k^{\dagger}D_k + \phi_k^{\dagger}\phi_k \right)
 -\hbar \delta \sum\limits_k    \left(
P_k^{\dagger} P_k + Q_k^{\dagger}Q_k  \right) 
-\hbar\Delta_{\theta}  \sum\limits_k
 \left( U_k^{\dagger} U_k +  V_k^{\dagger} V_k \right) \label{hr0} \\
 & + \hbar  \sum\limits_{k} \omega_k  
 \left(\phi_k^{\dagger}D_k + D_k^{\dagger} \phi_k \right) 
 -\hbar\Omega_0\sum\limits_k\left(P_k^{\dagger}\phi_k+P_k\phi_k^{\dagger} 
+Q_k^{\dagger}D_k + D_k^{\dagger}Q_k \right) \notag \\
& -\hbar\ve\sum\limits_{\genfrac{}{}{0pt}{2}{m}{m\not=0}}\sum\limits_k X_k^{\dagger}(m) X_k(m) 
  -\hbar\delta\sum\limits_{\genfrac{}{}{0pt}{2}{m}{m\not=0}}\sum\limits_k H_k^{\dagger}(m) H_k(m) 
 -\hbar\Delta_{\theta} \sum\limits_{\genfrac{}{}{0pt}{2}{m}{m\not=0}}\sum\limits_k I_k^{\dagger}(m) I_k(m) 
,\notag
\end{align}
\begin{align}
H_{\text{B}}^{(1)}=
&  -\frac{1}{2}\hbar\Omega_0 \cos\theta \sum\limits_k\left[
X_k(-2)\left(P_k^{\dagger} - Q_k^{\dagger}\right) 
 +  X_k(2) \left(P_k^{\dagger} + Q_k^{\dagger}\right) \right]  \label{hr1} \\
& - \frac{1}{\sqrt{2}}\hbar\Omega_0 \cos\theta \sum\limits_k\Big[
\sum\limits_{\genfrac{}{}{0pt}{2}{m}{m\not=0,-2}}
X_k(m)H_k^{\dagger}(m+1) 
+ \sum\limits_{\genfrac{}{}{0pt}{2}{m}{m\not=0,2}}
X_k(m)H_k^{\dagger}(m-1) \Big] + \text{h.c.}, \notag 
\end{align}
\begin{align}
H_{\text{B}}^{(2)}=
& - \hbar g_2 \cos\theta \sum\limits_{k} \Big[
U_k^{\dagger} \sum\limits_{p}
 \phi_p\phi_{k-p}
 +
V_k^{\dagger} \sum\limits_{p} 
D_p \phi_{k-p} \Big] + \text{h.c.}  , \label{hr2}  
\end{align}
\begin{align}
H_{\text{B}}^{(3)} = 
& -\frac{1}{2}\hbar g_2 \sum\limits_{k}\sum\limits_{p} \left[
\left(U_k^{\dagger}-V_k^{\dagger}\right)X_{k-p}(-2)
\left(\phi_p + D_p\right)
+\left(U_k^{\dagger}+V_k^{\dagger}\right)X_{k-p}(2)
\left(\phi_p - D_p\right)
  \right] \label{hr3} \\
& -\frac{1}{\sqrt{2}}\hbar g_2 \sum\limits_k\sum\limits_{p}\Big[
\sum\limits_{\genfrac{}{}{0pt}{2}{m}{m\not=0,-2}}
 I_k^{\dagger}(m+1) X_{k-p}(m)(\phi_p+D_p)
+  \sum\limits_{\genfrac{}{}{0pt}{2}{m}{m\not=0,2}}
 I_k^{\dagger}(m-1)  X_{k-p}(m)(\phi_p-D_p)\Big]
+\text{h.c.} . \notag  
\end{align}
The dominant contribution to the bath Hamiltonian $H_{\text{B}}$ is 
represented by $H_{\text{B}}^{(0)}$. The term $H_{\text{B}}^{(1)}$ accounts 
for the modification of stationary light due to the fast oscillating spin coherences. 
Furthermore, $H_{\text{B}}^{(2)}$ and $H_{\text{B}}^{(3)}$  arise from the coupling 
of bath excitations to the transition $\ket{4}\leftrightarrow\ket{2}$. 
We write the interaction Hamiltonian in Eq.~(\ref{lr}) as a sum of 
three parts 
\be
V = V^{(0)} + V^{(1)}+ V^{(2)}, \qquad \text{where}
\ee 
\begin{align}
 V^{(0)} = & -\hbar  \cos\theta \Delta\omega \sum\limits_{k} 
\left(\psi_k^{\dagger}\phi_k +\psi_k\phi_k^{\dagger} \right)   
 + \hbar \cos\theta \sum\limits_{k} \omega_k  
\left(\psi_k^{\dagger}D_k + D_k^{\dagger} \psi_k \right) \label{v0} \\
& + \hbar g_2 \cos\theta \sum\limits_{k}  \left(
U_k^{\dagger} \sum\limits_{p}
 \psi_p\psi_{k-p}+ U_k \sum\limits_{p}
 \psi_p^{\dagger} \psi_{k-p}^{\dagger} \right), \notag 
\end{align}
\begin{align}
 V^{(1)} = &   \hbar g_2   \sum\limits_{k} \Big(
U_k^{\dagger} \sum\limits_{p} \phi_p\psi_{k-p} +  
V_k^{\dagger} \sum\limits_{p}   D_p\psi_{k-p} \Big)  + \text{h.c.} ,
  \label{v1} 
\end{align}
\begin{align}
 V^{(2)} = 
& -\frac{1}{2}\hbar g_2 \cos\theta \sum\limits_{k} \sum\limits_{p}
\left[  \left(U_k^{\dagger} - V_k^{\dagger} \right) \psi_p X_{k-p}(-2)
 + \left(U_k^{\dagger} + V_k^{\dagger} \right) \psi_p X_{k-p}(2)
   \right] \label{v2} \\
& -\frac{1}{\sqrt{2}}\hbar g_2 \cos\theta 
\sum\limits_k\sum\limits_{p} \Big[
\sum\limits_{\genfrac{}{}{0pt}{2}{m}{m\not=0,-2}}
I_k^{\dagger}(m+1) \psi_p   X_{k-p}(m)
+\sum\limits_{\genfrac{}{}{0pt}{2}{m}{m\not=0,2}}
I_k^{\dagger}(m-1)\psi_p X_{k-p}(m) \Big]
+\text{h.c.} . 
\end{align}
The term $V^{(0)}$ describes the coupling of one or two dark-state polaritons to one 
bath excitation. On the other hand, $V^{(1)}$ arises from the coupling of one dark-state 
polariton and one bath excitation to the transition $\ket{4}\leftrightarrow\ket{2}$. 
The remaining part  $V^{(2)}$ accounts for the coupling of a dark-state polariton and a fast 
spin coherence to the transition $\ket{4}\leftrightarrow\ket{2}$. 
Finally, the decay term $\mc{L}_{\gamma}^{(\text{B})}\tilde{\vro}$ in Eq.~(\ref{lr}) reads
\begin{align}
\mc{L}_{\gamma}^{(\text{B})}\tilde{\vro} = & -\frac{\Gamma}{2} \sum\limits_k
\left( P_k^{\dagger}P_k \tilde{\vro} + \tilde{\vro} P_k^{\dagger}P_k -2 P_k \tilde{\vro} P_k^{\dagger} \right) 
 - \frac{\Gamma}{2} \sum\limits_k
\left( Q_k^{\dagger}Q_k \tilde{\vro} + \tilde{\vro} Q_k^{\dagger}Q_k -2 Q_k \tilde{\vro} Q_k^{\dagger} \right) \label{decay} \\
& - \frac{\Gamma}{2} \sum\limits_{\genfrac{}{}{0pt}{2}{m}{m\not= \pm 1}}\sum\limits_k
\left[ H_k^{\dagger}(m)H_k(m) \tilde{\vro}  
+ \tilde{\vro} H_k^{\dagger}(m)H_k(m) -2 H_k(m) \tilde{\vro} H_k^{\dagger}(m) \right] \notag \\
& -\frac{\gamma_{42}}{2}\sum\limits_k
\left( U_k^{\dagger}U_k \tilde{\vro} + \tilde{\vro} U_k^{\dagger}U_k -2 U_k \tilde{\vro} U_k^{\dagger} \right) 
-\frac{\gamma_{42}}{2}\sum\limits_k
\left( V_k^{\dagger}V_k \tilde{\vro} + \tilde{\vro} V_k^{\dagger}V_k -2 V_k \tilde{\vro} V_k^{\dagger} \right)  \notag\\
& -\frac{\gamma_{42}}{2} \sum\limits_{\genfrac{}{}{0pt}{2}{m}{m\not= \pm 1}}\sum\limits_k
\left[ I_k^{\dagger}(m)I_k(m) \tilde{\vro} 
+ \tilde{\vro} I_k^{\dagger}(m)I_k(m)   -2 I_k(m) \tilde{\vro} I_k^{\dagger}(m) \right] . 
\notag
\end{align}
\end{widetext}

\section{BATH CORRELATION FUNCTIONS \label{bath}}
Here we outline the calculation of the bath correlation functions 
\be
\text{Tr}_{\text{B}}\left\{B_i e^{\mc{L}_{\text{B}} \tau} B_j^{\dagger}\rf\right\} \label{cf}
\ee
that enter the master equation for dark-state polaritons 
via Eqs.~(\ref{master2}) and~(\ref{sint2}). 
In the following, we approximate the Liouvillian $\mc{L}_{\text{B}}$ by 
\be 
\mc{L}_{\text{B}}^{(0)} \rd = -\frac{i}{\hbar}[H_{\text{B}}^{(0)},\rd] 
+ \mc{L}_{\gamma}^{(\text{B})}\rd ,\label{lr2}
\ee
which amounts to neglect $H_{\text{B}}^{(1)}$, $H_{\text{B}}^{(2)}$ and $H_{\text{B}}^{(3)}$ 
in the expression for $H_{\text{B}}$. 
In a first step, we show how the bath dynamics can be solved exactly with respect to 
the super-operator in Eq.~(\ref{lr2}), and then we specify the conditions that allow us to 
neglect $H_{\text{B}}^{(1)}$, $H_{\text{B}}^{(2)}$ and $H_{\text{B}}^{(3)}$. 
The simplified bath dynamics according to Eq.~(\ref{lr2})  justifies to 
replace $V$ in the commutator $[V,\ldots]$ of Eq.~(\ref{sint}) 
by $V^{(-)}$, see Sec.~\ref{elemination}.  
Furthermore, we argue that the dominant contribution to Eq.~(\ref{sint2}) stems from the 
interaction Hamiltonian $V^{(0)}$, while $V^{(1)}$ and $V^{(2)}$ can be neglected. 
At the end of this section, we show that $V^{(1)}$ is indeed negligible if the 
dynamics of the system is restricted to the subspace $\hf$. 
In addition,  all correlation functions in Eq.~(\ref{cf}) 
where either $B_i$, $B_j$ or both stem from  $V^{(2)}$ vanish. 
 $V^{(2)}$ may thus only  contribute to higher-order terms beyond 
the Born approximation, but this effect will be small in the slow light limit since  
$V^{(2)}$ is proportional to $\cos\theta$. 

Here we only   take into account the bath operators $B_i \in\{ D_k,\phi_k,U_k\}$ 
that appear in the  interaction Hamiltonian $V_0$ in Eq.~(\ref{v0}). 
In principle, all combinations of these bath operators 
can enter the correlation functions in  Eq.~(\ref{sint2}).  
Their evaluation can be accomplished if 
the correlation functions in Eq.~(\ref{cf}) are  regarded 
as mean values of an operator $B_i$ with respect to the 
time-dependent, non-hermitian  operator 
$\hat{X}= e^{\mc{L}_{\text{B}} \tau} B_j^{\dagger}\rf$, 
\begin{align}
\meanX{B_i} = \text{Tr}_{\text{B}}\left\{B_i \hat{X} \right\}\,.
\end{align}
It follows that the equations of motion for these mean values are given by
\be 
\partial_t \meanX{B_i} = -\frac{i}{\hbar}\text{Tr}_{\text{B}}\left([B_i,H_{\text{B}}^{(0)}]\hat{X} \right) 
+ \text{Tr}_{\text{B}}\left(B_i \mc{L}_{\gamma}^{(\text{B})} \hat{X} \right) \,.
\ee
If we apply this result to the operator $U_k$, we find that 
the time evolution of $\meanX{U_k}$ is determined by  a single equation
\begin{align}
& \partial_t \meanX{U_k} 
= \left( i\Delta_{\theta} -\frac{\gamma_{42}}{2} \right) \meanX{U_k} \,. 
\label{dU}
\end{align}
On the other hand, the mean values of $\meanX{D_k}$ and $\meanX{\phi_k}$ 
are coupled to $\meanX{P_k}$ and $\meanX{Q_k}$ 
via the following set of linear equations,
\begin{align}
& \partial_t \meanX{\phi_k} 
= i \Delta\omega \meanX{\phi_k} - i \omega_k \meanX{D_k} + i \Omega_0 \meanX{P_k}
\,,\notag \\
& \partial_t \meanX{P_k} 
= \left( i \delta -\Gamma/2 \right) \meanX{P_k} 
+ i \Omega_0\meanX{\phi_k}   \,,\notag \\ 
&\partial_t \meanX{D_k} 
= i \Delta\omega \meanX{D_k} -i\omega_k \meanX{\phi_k} + i\Omega_0 \meanX{Q_k} \,,\notag \\ 
&\partial_t \meanX{Q_k} 
= \left( i\delta - \Gamma/2 \right) \meanX{Q_k} 
+ i \Omega_0  \meanX{D_k}\,.
\label{dsys}
\end{align}
It follows that the only non-vanishing terms in Eq.~(\ref{sint2}) are given by 
\begin{align}
\mc{S}(\rd) 
=& \Delta\omega^2  \cos^2\theta    \sum\limits_{k,p}  \mc{I}_1 
\left( \psi_k^{\dagger} \psi_p \rd - \psi_p \rd \psi_k^{\dagger}\right) \label{sint3}  \\
& + \cos^2\theta  \sum\limits_{k,p} \mc{I}_2 
 \omega_k \omega_p
\left( \psi_k^{\dagger} \psi_p \rd - \psi_p \rd \psi_k^{\dagger}\right) \notag\\
& + g_2^2 \cos^2\theta   \sum\limits_{k,p}
 \mc{I}_3  \sum\limits_{k^{\prime},p^{\prime}}\notag \\
& \times
 \left(  \psi_{k^{\prime}}^{\dagger} \psi_{k-k^{\prime}}^{\dagger}
\psi_{p^{\prime}}\psi_{p-p^{\prime}} \rd - \psi_{p^{\prime}}\psi_{p-p^{\prime}} \rd
 \psi_{k^{\prime}}^{\dagger}  \psi_{k-k^{\prime}}^{\dagger} \right) \notag \\
& - \cos^2\theta \Delta\omega \sum\limits_{k,p} \omega_k 
\mc{I}_4 
\left( \psi_k^{\dagger} \psi_p \rd - \psi_p \rd \psi_k^{\dagger}\right) \notag\\
& - \cos^2\theta \Delta\omega \sum\limits_{k,p} \omega_p 
\mc{I}_5 
\left( \psi_k^{\dagger} \psi_p \rd - \psi_p \rd \psi_k^{\dagger}\right)  ,\notag
\end{align}
where  the integrals over the bath correlation functions are defined as 
\begin{align}
\mc{I}_1(k,p) & = 
\int\limits_0^{\infty} d\tau 
\text{Tr}_{\text{B}}\left[\phi_k e^{\mc{L}_{\text{B}} \tau} \phi_p^{\dagger}\rf\right] \,,\label{i1}  \\
\mc{I}_2(k,p) & =  
\int\limits_0^{\infty} d\tau 
\text{Tr}_{\text{B}}\left[D_k e^{\mc{L}_{\text{B}} \tau} D_p^{\dagger}\rf\right]\,, \label{i2} \\
\mc{I}_3(k,p) & = \int\limits_0^{\infty} d\tau 
\text{Tr}_{\text{B}}\left[U_k e^{\mc{L}_{\text{B}} \tau} U_p^{\dagger}\rf\right] \,, \label{i3} \\
\mc{I}_4(k,p) & =  \int\limits_0^{t} d\tau 
\text{Tr}_{\text{B}}\left[D_k e^{\mc{L}_{\text{B}} \tau} \phi_p^{\dagger}\rf\right]\,, \label{i4} \\
\mc{I}_5(k,p) & = \int\limits_0^{t} d\tau 
\text{Tr}_{\text{B}}\left[\phi_k e^{\mc{L}_{\text{B}} \tau} D_p^{\dagger}\rf\right] \,. \label{i5}
\end{align}
We illustrate the evaluation of the integrals in Eqs.~(\ref{i1})-(\ref{i5}) using the example of 
\be
 \mc{I}_1    = \int\limits_0^{\infty} d\tau \meanX{\phi_k}\,,
\ee 
where the mean value $\meanX{\phi_k}$ is taken with respect to $\hat{X}= \phi_p^{\dagger} \rf$. 
The integral in the latter equation can be regarded as 
the Laplace transform of $\meanX{\phi_k}$ evaluated at $s=0$. 
In order to   determine $\mc{I}_1$, 
we write the system of differential equations~(\ref{dsys}) in matrix form, 
\begin{align}
 \partial_t \mf{y} = M \mf{y} \,,
\label{dy}
\end{align}
where $M$ is a $4\times 4$ matrix  and 
\be
\mf{y} = (\meanX{\phi_k},\meanX{P_k},\meanX{D_k},\meanX{Q_k}).
\ee
Since all mean values tend to zero for $t\rightarrow\infty$ due to 
the presence of the decay term $\mc{L}_{\gamma}^{(\text{B})} \rd$, the Laplace transform 
$\tilde{\mf{y}}(s)$ of $\mf{y}(t)$ exists and Eq.~(\ref{dy}) 
yields  $s \tilde{\mf{y}}(s) -\mf{y}(0) = M \tilde{\mf{y}}(s)$. 
In the limit $s\rightarrow 0$, we thus obtain 
\be 
\tilde{\mf{y}}(0) = -M^{-1} \mf{y}(0), \label{ylap}
\ee 
where $\mf{y}(0)$ represents $\mf{y}$ at time $t=0$. 
Since the mean values  are taken with respect 
to $\hat{X}= \phi_p^{\dagger}\rf$,  we have $\mf{y}(0) = (1,0,0,0)$ 
and thus $\tilde{\mf{y}}(0)$ can be determined. 
Finally, $\mc{I}_1$ can be identified with $[\tilde{\mf{y}}(0)]_1$, i.e., the 
first component of $\tilde{\mf{y}}(0)$ in Eq.~(\ref{ylap}). 
The evaluation of the remaining integrals follows the same route and yields 
\begin{align}
 &\text{Re}[\mc{I}_1] = \text{Re}[\mc{I}_2] 
\approx \frac{\Gamma}{2\Omega_0^2} \delta(k,p)\,, \label{c1}\\
&\text{Im}[\mc{I}_1] = \text{Im}[\mc{I}_2] 
\approx -\frac{\delta}{\Omega_0^2} \delta(k,p)  \,,\\
 &\text{Re}[\mc{I}_3] =  \frac{\gamma_{42}/2}{ \Delta_{\theta}^2 + \gamma_{42}^2/4}\delta(k,p)\,,\label{c2} \\
&\text{Im}[\mc{I}_3] = \frac{\Delta_{\theta}}{ \Delta_{\theta}^2 + \gamma_{42}^2/4}\delta(k,p)\,,  \label{c3}\\
& \mc{I}_4 = \mc{I}_5 \approx 0\,. \label{c4}
\end{align}
These simple expressions for the integrals represent an expansion of 
more complicated terms that holds if $\Omega_0$ is sufficiently large 
as compared to the detuning $|\delta|$ and the decay rates of the excited 
states [see Eq.~(\ref{approx1})]. In addition,  
$\Delta\omega$ and $|\omega_k|_{\text{max}}$ must be at most of the order of  $\Omega_0$. 
If the expressions in Eqs.~(\ref{c1})-(\ref{c4}) are plugged into Eq.~(\ref{sint3}), 
we obtain the final result for our master equation~(\ref{meq4}). 
The validity of the Markov approximation requires that the decay of the bath functions 
in Eqs.~(\ref{i1})-(\ref{i5}) is fast as compared to the change of the density operator 
introduced by these terms. Since the correlation functions decay on a timescale that is 
of the order of $1/\gamma_{ij}$, the Markov approximation is justified if 
the conditions in Eqs.~(\ref{markov1}) and (\ref{markov3}) are met. 

Next we specify the conditions that allow us to neglect 
$H_{\text{B}}^{(1)}$, $H_{\text{B}}^{(2)}$ and $H_{\text{B}}^{(3)}$. 
The Hamiltonian $H_{\text{B}}^{(1)}$ describes the modification of stationary 
light due to the fast oscillating  spin excitations $X_k(m)$ ($m\not=0$). 
These excitations are washed out due to the motion of the atoms, and the 
corresponding decay rate $\Gamma_{\text{FO}}$  depends on the temperature 
of the atomic cloud. Note that the decay of the slowly varying spin excitations $X_k(0)$ 
is significantly smaller than $\Gamma_{\text{FO}}$ since the relevant wavenumbers 
are several orders of magnitude smaller. 
We emphasize that a Markovian master equation for the 
dark-state polaritons corresponding to the level scheme in Fig.~\ref{fig2}  
is only possible if $\Gamma_{\text{FO}}$ is comparable to 
the decay rate of the excited states. 
More specifically,  the Hamiltonian $H_{\text{B}}^{(1)}$ alters the set of equations~(\ref{dsys}) 
and introduces a coupling between $\meanX{P_k}$, $\meanX{Q_k}$ and the fast spin coherences 
$\meanX{X_k(\pm 2)}$. 
Due to the decay of $\meanX{X_k(\pm 2)}$, the effective coupling between 
$\meanX{P_k}$, $\meanX{Q_k}$ and $\meanX{X_k(\pm 2)}$ is given by 
$(\Omega_0 \cos\theta)^2/\Gamma_{\text{FO}}$ if
$\Omega_0 \cos\theta\ll \Gamma_{\text{FO}}$~\cite{comment1}. 
It follows that the effect of $H_{\text{B}}^{(1)}$ is negligible provided that 
the condition in Eq.~(\ref{fcc}) holds, 
which means that the effective coupling 
between $\meanX{P_k}$, $\meanX{Q_k}$ and $\meanX{X_k(\pm 2)}$
is negligible on the timescale $1/\Gamma$ which represents the lifetime of 
excitations in modes $\{D_k,\phi_k,P_k,Q_k\}$. 
Note that the decay of fast oscillating coherences and condition~(\ref{fcc}) 
is not required in the case of the level scheme discussed 
in Sec.~\ref{other}. 

The Hamiltonian $H_{\text{B}}^{(2)}$ gives rise to a modification of 
Eq.~(\ref{dU}) that determines $\meanX{U_k}$,
\begin{align}
& \partial_t \meanX{U_k} 
= \left( i\Delta_{\theta} -\frac{\gamma_{42}}{2} \right) \meanX{U_k} 
+ i g_2  \cos\theta 
 \sum\limits_{p} \meanX{\phi_p \phi_{k-p}} \,. 
\label{dU2}
\end{align}
If the probe field modes form a (quasi-)continuum, then the second term in 
Eq.~(\ref{dU2}) gives rise to an additional decay channel of excitations 
in the mode $U_k$. We find that the associated decay rate is at most 
given by 
\be
\Gamma_{U}=\Gamma_{\text{1D}} \Omega_0\cos^2\theta/\Gamma, 
\ee
where $\Gamma_{\text{1D}}= g_2^2L/c$ is the decay rate of the excited 
state $\ket{4}$ into the fiber modes. It follows that 
the influence of $H_{\text{B}}^{(2)}$ is negligible provided that  
$\Gamma_{U}$ is much smaller than $\gamma_{42}$, which can always be achieved 
in the slow-light regime.

It remains to discuss the impact of $H_{\text{B}}^{(3)}$ and $V^{(1)}$. 
These terms are negligible since the physical processes described by them 
are off-resonant and therefore strongly suppressed. 
Formally, the latter result can be obtained via a rotating-wave type approximation if  
the operators $\phi_k$ and $D_k$ in $H_{\text{B}}^{(3)}$ and $V^{(1)}$ are expressed in terms of 
new bosonic operators that diagonalize $H_{\text{B}}^{(0)}$. 
The influence of these operators is found to be small 
if $g_2/\Omega_0\propto 1/N \ll1$, where $N$ is the total number of atoms.  

Finally, we note that we have verified the validity of the approximations discussed above 
by the numerical comparison of our master equation  with the results of
the full dynamics for a single mode. 

\section{MAXWELL-BLOCH EQUATIONS \label{mbe}} 
Here we outline the numerical integration of the coupled Maxwell-Bloch 
equations~\cite{zubairy:qo} for classical probe and control 
fields that interact with the $\Lambda$ subsystem formed by states $\ket{1}$, $\ket{2}$ and $\ket{3}$. 
The density operator  of a 
single atom at position $z$ is denoted by $R$, and the coherence 
$R_{31}=\bra{3}R\ket{1}$ is written as the sum of two counter-propagating terms, 
\begin{align}
& R_{31}(z,t)= R_{31}^{(+)}(z,t) e^{i k_p z} + R_{31}^{(-)}(z,t) e^{-i k_p z}. 
\end{align}
We assume that the probe fields are weak such that we can set $R_{11}\approx 1$, 
and  apply the secular approximation where we drop fast oscillating terms $\exp[\pm 2 i k_p z]$. 
The Bloch equations of the system are thus given by~\cite{zimmer:06,kiffner:09}  
\begin{align}
& \partial_t R_{31}^{(\pm)} = i \Omega_c e^{\mp i \frac{\Delta\omega}{c} z } R_{21} 
+ (i\delta - \Gamma/2) R_{31}^{(+)}  + i \mc{G}_{\pm} \\
& \partial_t R_{21} = i \ve R_{21} +i \Omega_c
\left( e^{i\frac{\Delta\omega}{c} z}R_{31}^{(+)} +e^{-i\frac{\Delta\omega}{c} z}R_{31}^{(-)} \right),
\label{bloch}
\end{align}
where $\mc{G}_+$ ($\mc{G}_-$) is the Rabi frequency of the classical probe field that propagates 
in the positive (negative) $z$ direction, and $k_p=\omega_p/c$ is the wave number corresponding 
to the central frequency $\omega_p$ of the probe field. Note that $\mc{G}_{\pm}(z,t)$ are 
slowly varying functions of position and time. On the other hand, the Rabi frequency $\Omega_c$ 
of the each control field is assumed to be position-independent but varies with time.  
The Bloch equations have to be solved consistently with Maxwell's equations that 
yield~\cite{zimmer:06,kiffner:09}
\begin{align}
 & \left( \frac{1}{c}\partial_t \pm \partial_z \right) \mc{G}_{\pm} = i \frac{N g_1^2}{c} R_{31}^{(\pm)}, 
\label{max} 
\end{align}
where we employed the slowly varying envelope approximation~\cite{zubairy:qo}. 
The set of equations~(\ref{bloch})  and~(\ref{max})  allows us  to determine 
$\mc{G}_{\pm}$ as well as the atomic variables $R$. 
Note that equivalent results without the secular approximation can be obtained~\cite{zimmer:08} if the 
level scheme in Fig.~\ref{fig4} is employed.

In the slow light limit, the expectation value of the polariton pulse 
in position space is directly proportional to the expectation value 
of the ground-state coherence, 
\be
\mean{\psi(z)} \propto \mean{R_{12}}(z). 
\ee
It follows that $\mean{\psi_k}$ can be calculated from $\mean{R_{12}}(z)$ 
via a Fourier transformation with respect to position,  and we have 
$\mean{\psi_k^{\dagger}\psi_k}=|\mean{\psi_k}|^2$ 
since a classical probe field corresponds to a coherent state. 



\begin{thebibliography}{60}
\expandafter\ifx\csname natexlab\endcsname\relax\def\natexlab#1{#1}\fi
\expandafter\ifx\csname bibnamefont\endcsname\relax
  \def\bibnamefont#1{#1}\fi
\expandafter\ifx\csname bibfnamefont\endcsname\relax
  \def\bibfnamefont#1{#1}\fi
\expandafter\ifx\csname citenamefont\endcsname\relax
  \def\citenamefont#1{#1}\fi
\expandafter\ifx\csname url\endcsname\relax
  \def\url#1{\texttt{#1}}\fi
\expandafter\ifx\csname urlprefix\endcsname\relax\def\urlprefix{URL }\fi
\providecommand{\bibinfo}[2]{#2}
\providecommand{\eprint}[2][]{\url{#2}}

\bibitem[{\citenamefont{Nielsen and Chuang}(2000)}]{nielsen:00}
\bibinfo{author}{\bibfnamefont{M.~A.} \bibnamefont{Nielsen}} \bibnamefont{and}
  \bibinfo{author}{\bibfnamefont{I.~L.} \bibnamefont{Chuang}},
  \emph{\bibinfo{title}{Quantum Computation and Quantum Information}}
  (\bibinfo{publisher}{Cambridge University Press},
  \bibinfo{address}{Cambridge}, \bibinfo{year}{2000}).

\bibitem[{\citenamefont{Bloch et~al.}(2008)\citenamefont{Bloch, Dalibard, and
  Zwerger}}]{bloch:08}
\bibinfo{author}{\bibfnamefont{I.}~\bibnamefont{Bloch}},
  \bibinfo{author}{\bibfnamefont{J.}~\bibnamefont{Dalibard}}, \bibnamefont{and}
  \bibinfo{author}{\bibfnamefont{W.}~\bibnamefont{Zwerger}},
  \bibinfo{journal}{Rev. Mod. Phys.} \textbf{\bibinfo{volume}{80}},
  \bibinfo{pages}{885} (\bibinfo{year}{2008}).

\bibitem[{\citenamefont{Hartmann et~al.}(2006)\citenamefont{Hartmann,
  {{Brand\~{a}o}}, and Plenio}}]{HBP06}
\bibinfo{author}{\bibfnamefont{M.~J.} \bibnamefont{Hartmann}},
  \bibinfo{author}{\bibfnamefont{F.~G. S.~L.} \bibnamefont{{{Brand\~{a}o}}}},
  \bibnamefont{and} \bibinfo{author}{\bibfnamefont{M.~B.}
  \bibnamefont{Plenio}}, \bibinfo{journal}{Nat. Phys.}
  \textbf{\bibinfo{volume}{2}}, \bibinfo{pages}{849} (\bibinfo{year}{2006}).

\bibitem[{\citenamefont{Hartmann
  et~al.}(2008{\natexlab{a}})\citenamefont{Hartmann, {Brand\~{a}o}, and
  Plenio}}]{HBP08}
\bibinfo{author}{\bibfnamefont{M.~J.} \bibnamefont{Hartmann}},
  \bibinfo{author}{\bibfnamefont{F.~G. S.~L.} \bibnamefont{{Brand\~{a}o}}},
  \bibnamefont{and} \bibinfo{author}{\bibfnamefont{M.~B.}
  \bibnamefont{Plenio}}, \bibinfo{journal}{Laser \& Photon. Rev.}
  \textbf{\bibinfo{volume}{2}}, \bibinfo{pages}{527}
  (\bibinfo{year}{2008}{\natexlab{a}}).

\bibitem[{\citenamefont{Rossini and Fazio}(2007)}]{RF07}
\bibinfo{author}{\bibfnamefont{D.}~\bibnamefont{Rossini}} \bibnamefont{and}
  \bibinfo{author}{\bibfnamefont{R.}~\bibnamefont{Fazio}},
  \bibinfo{journal}{Phys. Rev. Lett.} \textbf{\bibinfo{volume}{99}},
  \bibinfo{pages}{186401} (\bibinfo{year}{2007}).

\bibitem[{\citenamefont{Angelakis et~al.}(2007)\citenamefont{Angelakis, Santos,
  and Bose}}]{ASB06}
\bibinfo{author}{\bibfnamefont{D. G.}~\bibnamefont{Angelakis}},
  \bibinfo{author}{\bibfnamefont{M.~F.} \bibnamefont{Santos}},
  \bibnamefont{and} \bibinfo{author}{\bibfnamefont{S.}~\bibnamefont{Bose}},
  \bibinfo{journal}{Phys. Rev. A} \textbf{\bibinfo{volume}{76}},
  \bibinfo{pages}{031805(R)} (\bibinfo{year}{2007}).

\bibitem[{\citenamefont{D.Gerace et~al.}(2006)\citenamefont{D.Gerace, T\"ureci,
  Imamo\v{g}lu, Giovannetti, and Fazio}}]{GTI+09}
\bibinfo{author}{\bibnamefont{D.Gerace}}, \bibinfo{author}{\bibfnamefont{H.~E.}
  \bibnamefont{T\"ureci}},
  \bibinfo{author}{\bibfnamefont{A.}~\bibnamefont{Imamo\v{g}lu}},
  \bibinfo{author}{\bibfnamefont{V.}~\bibnamefont{Giovannetti}},
  \bibnamefont{and} \bibinfo{author}{\bibfnamefont{R.}~\bibnamefont{Fazio}},
  \bibinfo{journal}{Nat. Phys.} \textbf{\bibinfo{volume}{5}},
  \bibinfo{pages}{281} (\bibinfo{year}{2006}).

\bibitem[{haf({\natexlab{a}})}]{hafezi:09a}
\bibinfo{note}{M. Hafezi and D. E. Chang and V. Gritsev and E. Demler and M.
  Lukin, arXiv:0907.5206 (2009).}

\bibitem[{haf({\natexlab{b}})}]{hafezi:09}
\bibinfo{note}{M. Hafezi and D. E. Chang and V. Gritsev and E. Demler and M.
  Lukin, arXiv:0911.4766 (2009).}

\bibitem[{\citenamefont{Carusotto et~al.}(2009)\citenamefont{Carusotto, Gerace,
  Tureci, De~Liberato, Ciuti, and Imamo\v{g}lu}}]{carusotto:09}
\bibinfo{author}{\bibfnamefont{I.}~\bibnamefont{Carusotto}},
  \bibinfo{author}{\bibfnamefont{D.}~\bibnamefont{Gerace}},
  \bibinfo{author}{\bibfnamefont{H.~E.} \bibnamefont{Tureci}},
  \bibinfo{author}{\bibfnamefont{S.}~\bibnamefont{De~Liberato}},
  \bibinfo{author}{\bibfnamefont{C.}~\bibnamefont{Ciuti}}, \bibnamefont{and}
  \bibinfo{author}{\bibfnamefont{A.}~\bibnamefont{Imamo\v{g}lu}}, \bibinfo{journal}{Phys. Rev. Lett.}
  \textbf{\bibinfo{volume}{103}}, \bibinfo{pages}{033601}
  (\bibinfo{year}{2009}).

\bibitem[{\citenamefont{Fleischhauer et~al.}(2008)\citenamefont{Fleischhauer,
  Otterbach, and Unanyan}}]{fleischhauer:08}
\bibinfo{author}{\bibfnamefont{M.}~\bibnamefont{Fleischhauer}},
  \bibinfo{author}{\bibfnamefont{J.}~\bibnamefont{Otterbach}},
  \bibnamefont{and} \bibinfo{author}{\bibfnamefont{R.~G.}
  \bibnamefont{Unanyan}}, \bibinfo{journal}{Phys. Rev. Lett.}
  \textbf{\bibinfo{volume}{101}}, \bibinfo{pages}{163601}
  (\bibinfo{year}{2008}).

\bibitem[{\citenamefont{Chang et~al.}(2008)\citenamefont{Chang, Gritsev,
  Morigi, Vuleti\'{c}, Lukin, and Demler}}]{chang:08}
\bibinfo{author}{\bibfnamefont{D.~E.} \bibnamefont{Chang}},
  \bibinfo{author}{\bibfnamefont{V.}~\bibnamefont{Gritsev}},
  \bibinfo{author}{\bibfnamefont{G.}~\bibnamefont{Morigi}},
  \bibinfo{author}{\bibfnamefont{V.}~\bibnamefont{Vuleti\'{c}}},
  \bibinfo{author}{\bibfnamefont{M.~D.} \bibnamefont{Lukin}}, \bibnamefont{and}
  \bibinfo{author}{\bibfnamefont{E.~A.} \bibnamefont{Demler}},
  \bibinfo{journal}{Nat. Phys.} \textbf{\bibinfo{volume}{4}},
  \bibinfo{pages}{884} (\bibinfo{year}{2008}).

\bibitem[{\citenamefont{Kiffner and Hartmann}(2010)}]{kiffner:10}
\bibinfo{author}{\bibfnamefont{M.}~\bibnamefont{Kiffner}} \bibnamefont{and}
  \bibinfo{author}{\bibfnamefont{M.~J.} \bibnamefont{Hartmann}},
  \bibinfo{journal}{Phys. Rev. A} \textbf{\bibinfo{volume}{81}},
  \bibinfo{pages}{021806(R)} (\bibinfo{year}{2010}).

\bibitem[{\citenamefont{Hartmann
  et~al.}(2008{\natexlab{b}})\citenamefont{Hartmann, {Brand\~{a}o}, and
  Plenio}}]{hartmann:08}
\bibinfo{author}{\bibfnamefont{M.~J.} \bibnamefont{Hartmann}},
  \bibinfo{author}{\bibfnamefont{F.~G. S.~L.} \bibnamefont{{Brand\~{a}o}}},
  \bibnamefont{and} \bibinfo{author}{\bibfnamefont{M.~B.}
  \bibnamefont{Plenio}}, \bibinfo{journal}{New. J. Phys.}
  \textbf{\bibinfo{volume}{10}}, \bibinfo{pages}{033011}
  (\bibinfo{year}{2008}{\natexlab{b}}).

\bibitem[{\citenamefont{Hartmann}(2010)}]{hartmann:10}
\bibinfo{author}{\bibfnamefont{M.~J.} \bibnamefont{Hartmann}},
  \bibinfo{journal}{Phys. Rev. Lett.} \textbf{\bibinfo{volume}{104}},
  \bibinfo{pages}{113601} (\bibinfo{year}{2010}).

\bibitem[{\citenamefont{Fleischhauer and Lukin}(2000)}]{fleischhauer:00}
\bibinfo{author}{\bibfnamefont{M.}~\bibnamefont{Fleischhauer}}
  \bibnamefont{and} \bibinfo{author}{\bibfnamefont{M.~D.} \bibnamefont{Lukin}},
  \bibinfo{journal}{Phys. Rev. Lett.} \textbf{\bibinfo{volume}{84}},
  \bibinfo{pages}{5094} (\bibinfo{year}{2000}).

\bibitem[{\citenamefont{Fleischhauer and Lukin}(2002)}]{fleischhauer:02}
\bibinfo{author}{\bibfnamefont{M.}~\bibnamefont{Fleischhauer}}
  \bibnamefont{and} \bibinfo{author}{\bibfnamefont{M.~D.} \bibnamefont{Lukin}},
  \bibinfo{journal}{Phys. Rev. A} \textbf{\bibinfo{volume}{65}},
  \bibinfo{pages}{022314} (\bibinfo{year}{2002}).

\bibitem[{\citenamefont{Lukin}(2003)}]{lukin:03}
\bibinfo{author}{\bibfnamefont{M.~D.} \bibnamefont{Lukin}},
  \bibinfo{journal}{Rev. Mod. Phys.} \textbf{\bibinfo{volume}{75}},
  \bibinfo{pages}{457} (\bibinfo{year}{2003}).

\bibitem[{\citenamefont{Fleischhauer et~al.}(2005)\citenamefont{Fleischhauer,
  Imamo\v{g}lu, and Marangos}}]{fleischhauer:05}
\bibinfo{author}{\bibfnamefont{M.}~\bibnamefont{Fleischhauer}},
  \bibinfo{author}{\bibfnamefont{A.}~\bibnamefont{Imamo\v{g}lu}},
  \bibnamefont{and} \bibinfo{author}{\bibfnamefont{J.~P.}
  \bibnamefont{Marangos}}, \bibinfo{journal}{Rev. Mod. Phys.}
  \textbf{\bibinfo{volume}{77}}, \bibinfo{pages}{633} (\bibinfo{year}{2005}).

\bibitem[{\citenamefont{Hau et~al.}(1999)\citenamefont{Hau, Harris, Dutton, and
  Behroozi}}]{hau:99}
\bibinfo{author}{\bibfnamefont{L.~V.} \bibnamefont{Hau}},
  \bibinfo{author}{\bibfnamefont{S.~E.} \bibnamefont{Harris}},
  \bibinfo{author}{\bibfnamefont{Z.}~\bibnamefont{Dutton}}, \bibnamefont{and}
  \bibinfo{author}{\bibfnamefont{C.}~\bibnamefont{Behroozi}},
  \bibinfo{journal}{Nature} \textbf{\bibinfo{volume}{397}},
  \bibinfo{pages}{594} (\bibinfo{year}{1999}).

\bibitem[{\citenamefont{Kash et~al.}(1999)\citenamefont{Kash, Sautenkov,
  Zibrov, Hollberg, Welch, Lukin, Rostovtsev, Fry, and Scully}}]{kash:99}
\bibinfo{author}{\bibfnamefont{M.~M.} \bibnamefont{Kash}},
  \bibinfo{author}{\bibfnamefont{V.~A.} \bibnamefont{Sautenkov}},
  \bibinfo{author}{\bibfnamefont{A.~S.} \bibnamefont{Zibrov}},
  \bibinfo{author}{\bibfnamefont{L.}~\bibnamefont{Hollberg}},
  \bibinfo{author}{\bibfnamefont{G.~R.} \bibnamefont{Welch}},
  \bibinfo{author}{\bibfnamefont{M.~D.} \bibnamefont{Lukin}},
  \bibinfo{author}{\bibfnamefont{Y.}~\bibnamefont{Rostovtsev}},
  \bibinfo{author}{\bibfnamefont{E.~S.} \bibnamefont{Fry}}, \bibnamefont{and}
  \bibinfo{author}{\bibfnamefont{M. O.}~\bibnamefont{Scully}},
  \bibinfo{journal}{Phys. Rev. Lett.} \textbf{\bibinfo{volume}{82}},
  \bibinfo{pages}{5229} (\bibinfo{year}{1999}).

\bibitem[{\citenamefont{Budker et~al.}(1999)\citenamefont{Budker, Kimball,
  Rochester, and Yashchuk}}]{budker:99}
\bibinfo{author}{\bibfnamefont{D.}~\bibnamefont{Budker}},
  \bibinfo{author}{\bibfnamefont{D.~F.} \bibnamefont{Kimball}},
  \bibinfo{author}{\bibfnamefont{S.~M.} \bibnamefont{Rochester}},
  \bibnamefont{and} \bibinfo{author}{\bibfnamefont{V.~V.}
  \bibnamefont{Yashchuk}}, \bibinfo{journal}{Phys. Rev. Lett.}
  \textbf{\bibinfo{volume}{83}}, \bibinfo{pages}{1767} (\bibinfo{year}{1999}).

\bibitem[{\citenamefont{Liu et~al.}(2001{\natexlab{a}})\citenamefont{Liu,
  Dutton, Behroozi, and Hau}}]{hau:01}
\bibinfo{author}{\bibfnamefont{C.}~\bibnamefont{Liu}},
  \bibinfo{author}{\bibfnamefont{Z.}~\bibnamefont{Dutton}},
  \bibinfo{author}{\bibfnamefont{C.~H.} \bibnamefont{Behroozi}},
  \bibnamefont{and} \bibinfo{author}{\bibfnamefont{L.~V.} \bibnamefont{Hau}},
  \bibinfo{journal}{Nature} \textbf{\bibinfo{volume}{409}},
  \bibinfo{pages}{490} (\bibinfo{year}{2001}{\natexlab{a}}).

\bibitem[{\citenamefont{Phillips et~al.}(2001)\citenamefont{Phillips,
  Fleischhauer, Mair, Walsworth, and Lukin}}]{phillips:01}
\bibinfo{author}{\bibfnamefont{D.~F.} \bibnamefont{Phillips}},
  \bibinfo{author}{\bibfnamefont{A.}~\bibnamefont{Fleischhauer}},
  \bibinfo{author}{\bibfnamefont{A.}~\bibnamefont{Mair}},
  \bibinfo{author}{\bibfnamefont{R.~L.} \bibnamefont{Walsworth}},
  \bibnamefont{and} \bibinfo{author}{\bibfnamefont{M.~D.} \bibnamefont{Lukin}},
  \bibinfo{journal}{Phys. Rev. Lett.} \textbf{\bibinfo{volume}{86}},
  \bibinfo{pages}{783} (\bibinfo{year}{2001}).

\bibitem[{\citenamefont{Dey and Agarwal}(2003)}]{dey:03}
\bibinfo{author}{\bibfnamefont{T.~N.} \bibnamefont{Dey}} \bibnamefont{and}
  \bibinfo{author}{\bibfnamefont{G.~S.} \bibnamefont{Agarwal}},
  \bibinfo{journal}{Phys. Rev. A} \textbf{\bibinfo{volume}{67}},
  \bibinfo{pages}{033813} (\bibinfo{year}{2003}).

\bibitem[{\citenamefont{Andr\'{e} and Lukin}(2002)}]{andre:02}
\bibinfo{author}{\bibfnamefont{A.}~\bibnamefont{Andr\'{e}}} \bibnamefont{and}
  \bibinfo{author}{\bibfnamefont{M.~D.} \bibnamefont{Lukin}},
  \bibinfo{journal}{Phys. Rev. Lett.} \textbf{\bibinfo{volume}{89}},
  \bibinfo{pages}{143602} (\bibinfo{year}{2002}).

\bibitem[{\citenamefont{Bajcsy et~al.}(2003)\citenamefont{Bajcsy, Zibrov, and
  Lukin}}]{bajcsy:03}
\bibinfo{author}{\bibfnamefont{M.}~\bibnamefont{Bajcsy}},
  \bibinfo{author}{\bibfnamefont{A.~S.} \bibnamefont{Zibrov}},
  \bibnamefont{and} \bibinfo{author}{\bibfnamefont{M.~D.} \bibnamefont{Lukin}},
  \bibinfo{journal}{Nature} \textbf{\bibinfo{volume}{426}},
  \bibinfo{pages}{638} (\bibinfo{year}{2003}).

\bibitem[{\citenamefont{Zimmer et~al.}(2006)\citenamefont{Zimmer, Andr\'{e},
  Lukin, and Fleischhauer}}]{zimmer:06}
\bibinfo{author}{\bibfnamefont{F.~E.} \bibnamefont{Zimmer}},
  \bibinfo{author}{\bibfnamefont{A.}~\bibnamefont{Andr\'{e}}},
  \bibinfo{author}{\bibfnamefont{M.~D.} \bibnamefont{Lukin}}, \bibnamefont{and}
  \bibinfo{author}{\bibfnamefont{M.}~\bibnamefont{Fleischhauer}},
  \bibinfo{journal}{Opt. Commun.} \textbf{\bibinfo{volume}{264}},
  \bibinfo{pages}{441} (\bibinfo{year}{2006}).

\bibitem[{\citenamefont{Zimmer et~al.}(2008)\citenamefont{Zimmer, Otterbach,
  Unanyan, Shore, and Fleischhauer}}]{zimmer:08}
\bibinfo{author}{\bibfnamefont{F.~E.} \bibnamefont{Zimmer}},
  \bibinfo{author}{\bibfnamefont{J.}~\bibnamefont{Otterbach}},
  \bibinfo{author}{\bibfnamefont{R.~G.} \bibnamefont{Unanyan}},
  \bibinfo{author}{\bibfnamefont{B.~W.} \bibnamefont{Shore}}, \bibnamefont{and}
  \bibinfo{author}{\bibfnamefont{M.}~\bibnamefont{Fleischhauer}},
  \bibinfo{journal}{Phys. Rev. A} \textbf{\bibinfo{volume}{77}},
  \bibinfo{pages}{063823} (\bibinfo{year}{2008}).

\bibitem[{\citenamefont{Moiseev and Ham}(2005)}]{moiseev:05}
\bibinfo{author}{\bibfnamefont{S.~A.} \bibnamefont{Moiseev}} \bibnamefont{and}
  \bibinfo{author}{\bibfnamefont{B.~S.} \bibnamefont{Ham}},
  \bibinfo{journal}{Phys. Rev. A} \textbf{\bibinfo{volume}{71}},
  \bibinfo{pages}{053802} (\bibinfo{year}{2005}).

\bibitem[{\citenamefont{Moiseev and Ham}(2006)}]{moiseev:06}
\bibinfo{author}{\bibfnamefont{S.~A.} \bibnamefont{Moiseev}} \bibnamefont{and}
  \bibinfo{author}{\bibfnamefont{B.~S.} \bibnamefont{Ham}},
  \bibinfo{journal}{Phys. Rev. A} \textbf{\bibinfo{volume}{73}},
  \bibinfo{pages}{033812} (\bibinfo{year}{2006}).

\bibitem[{\citenamefont{Lin et~al.}(2009)\citenamefont{Lin, Liao, Peters, Chou,
  Wang, Cho, Kuan, and Yu}}]{lin:09}
\bibinfo{author}{\bibfnamefont{Y.-W.} \bibnamefont{Lin}},
  \bibinfo{author}{\bibfnamefont{W.-T.} \bibnamefont{Liao}},
  \bibinfo{author}{\bibfnamefont{T.}~\bibnamefont{Peters}},
  \bibinfo{author}{\bibfnamefont{H.-C.} \bibnamefont{Chou}},
  \bibinfo{author}{\bibfnamefont{J.-S.} \bibnamefont{Wang}},
  \bibinfo{author}{\bibfnamefont{H.-W.} \bibnamefont{Cho}},
  \bibinfo{author}{\bibfnamefont{P.-C.} \bibnamefont{Kuan}}, \bibnamefont{and}
  \bibinfo{author}{\bibfnamefont{I.~A.} \bibnamefont{Yu}},
  \bibinfo{journal}{Phys. Rev. Lett.} \textbf{\bibinfo{volume}{102}},
  \bibinfo{pages}{213601} (\bibinfo{year}{2009}).

\bibitem[{\citenamefont{Nikoghosyan and Fleischhauer}(2009)}]{nikoghosyan:09}
\bibinfo{author}{\bibfnamefont{G.}~\bibnamefont{Nikoghosyan}} \bibnamefont{and}
  \bibinfo{author}{\bibfnamefont{M.}~\bibnamefont{Fleischhauer}},
  \bibinfo{journal}{Phys. Rev. A} \textbf{\bibinfo{volume}{80}},
  \bibinfo{pages}{013818} (\bibinfo{year}{2009}).

\bibitem[{\citenamefont{Schmidt and Imamo\v{g}lu}(1996)}]{schmidt:96}
\bibinfo{author}{\bibfnamefont{H.}~\bibnamefont{Schmidt}} \bibnamefont{and}
  \bibinfo{author}{\bibfnamefont{A.}~\bibnamefont{Imamo\v{g}lu}},
  \bibinfo{journal}{Opt. Lett.} \textbf{\bibinfo{volume}{21}},
  \bibinfo{pages}{1936} (\bibinfo{year}{1996}).

\bibitem[{\citenamefont{Imamo\v{g}lu et~al.}(1997)\citenamefont{Imamo\v{g}lu,
  Schmidt, Woods, and Deutsch}}]{ISWD97}
\bibinfo{author}{\bibfnamefont{A.}~\bibnamefont{Imamo\v{g}lu}},
  \bibinfo{author}{\bibfnamefont{H.}~\bibnamefont{Schmidt}},
  \bibinfo{author}{\bibfnamefont{G.}~\bibnamefont{Woods}}, \bibnamefont{and}
  \bibinfo{author}{\bibfnamefont{M.}~\bibnamefont{Deutsch}},
  \bibinfo{journal}{Phys. Rev. Lett.} \textbf{\bibinfo{volume}{79}},
  \bibinfo{pages}{1467} (\bibinfo{year}{1997}).

\bibitem[{\citenamefont{Harris and Yamamoto}(1998)}]{harris:98}
\bibinfo{author}{\bibfnamefont{S.~E.} \bibnamefont{Harris}} \bibnamefont{and}
  \bibinfo{author}{\bibfnamefont{Y.}~\bibnamefont{Yamamoto}},
  \bibinfo{journal}{Phys. Rev. Lett.} \textbf{\bibinfo{volume}{81}},
  \bibinfo{pages}{3611} (\bibinfo{year}{1998}).

\bibitem[{\citenamefont{Harris and Hau}(1998)}]{harris:99}
\bibinfo{author}{\bibfnamefont{S.~E.} \bibnamefont{Harris}} \bibnamefont{and}
  \bibinfo{author}{\bibfnamefont{L. V.}~\bibnamefont{Hau}},
  \bibinfo{journal}{Phys. Rev. Lett.} \textbf{\bibinfo{volume}{82}},
  \bibinfo{pages}{4611} (\bibinfo{year}{1999}).

\bibitem[{\citenamefont{Kang and Zhu}(2003)}]{kang:03}
\bibinfo{author}{\bibfnamefont{H.}~\bibnamefont{Kang}} \bibnamefont{and}
  \bibinfo{author}{\bibfnamefont{Y.}~\bibnamefont{Zhu}},
  \bibinfo{journal}{Phys. Rev. Lett.} \textbf{\bibinfo{volume}{91}},
  \bibinfo{pages}{093601} (\bibinfo{year}{2003}).

\bibitem[{\citenamefont{Braje et~al.}(2004)\citenamefont{Braje, Balic, Goda,
  Yin, and Harris}}]{braje:04}
\bibinfo{author}{\bibfnamefont{D.~A.} \bibnamefont{Braje}},
  \bibinfo{author}{\bibfnamefont{V.}~\bibnamefont{Balic}},
  \bibinfo{author}{\bibfnamefont{S.}~\bibnamefont{Goda}},
  \bibinfo{author}{\bibfnamefont{G.~Y.} \bibnamefont{Yin}}, \bibnamefont{and}
  \bibinfo{author}{\bibfnamefont{S.~E.} \bibnamefont{Harris}},
  \bibinfo{journal}{Phys. Rev. Lett.} \textbf{\bibinfo{volume}{93}},
  \bibinfo{pages}{183601} (\bibinfo{year}{2004}).

\bibitem[{\citenamefont{Hartmann and Plenio}(2007)}]{hartmann:07}
\bibinfo{author}{\bibfnamefont{M.~J.} \bibnamefont{Hartmann}} \bibnamefont{and}
  \bibinfo{author}{\bibfnamefont{M.~B.} \bibnamefont{Plenio}},
  \bibinfo{journal}{Phys. Rev. Lett.} \textbf{\bibinfo{volume}{99}},
  \bibinfo{pages}{103601} (\bibinfo{year}{2007}).

\bibitem[{\citenamefont{Andr\'{e} et~al.}(2005)\citenamefont{Andr\'{e}, Bajcsy,
  Zibrov, and Lukin}}]{andre:05}
\bibinfo{author}{\bibfnamefont{A.}~\bibnamefont{Andr\'{e}}},
  \bibinfo{author}{\bibfnamefont{M.}~\bibnamefont{Bajcsy}},
  \bibinfo{author}{\bibfnamefont{A.~S.} \bibnamefont{Zibrov}},
  \bibnamefont{and} \bibinfo{author}{\bibfnamefont{M.~D.} \bibnamefont{Lukin}},
  \bibinfo{journal}{Phys. Rev. Lett.} \textbf{\bibinfo{volume}{94}},
  \bibinfo{pages}{063902} (\bibinfo{year}{2005}).

\bibitem[{\citenamefont{Aoki et~al.}(2006)\citenamefont{Aoki, Dayan, Wilcut,
  W.P.Bowen, Parkins, Kimble, Kippenberg, and Vahala}}]{aoki:06}
\bibinfo{author}{\bibfnamefont{T.}~\bibnamefont{Aoki}},
  \bibinfo{author}{\bibfnamefont{B.}~\bibnamefont{Dayan}},
  \bibinfo{author}{\bibfnamefont{E.}~\bibnamefont{Wilcut}},
  \bibinfo{author}{\bibnamefont{W.P.Bowen}},
  \bibinfo{author}{\bibfnamefont{A.}~\bibnamefont{Parkins}},
  \bibinfo{author}{\bibfnamefont{H.}~\bibnamefont{Kimble}},
  \bibinfo{author}{\bibfnamefont{T.}~\bibnamefont{Kippenberg}},
  \bibnamefont{and} \bibinfo{author}{\bibfnamefont{K.}~\bibnamefont{Vahala}},
  \bibinfo{journal}{Nature} \textbf{\bibinfo{volume}{443}},
  \bibinfo{pages}{671} (\bibinfo{year}{2006}).

\bibitem[{\citenamefont{Wallraff et~al.}(2004)\citenamefont{Wallraff, Schuster,
  Blais, Frunzio, Huang, Majer, Kumar, Girvin, and Schoelkopf}}]{wallraff:04}
\bibinfo{author}{\bibfnamefont{A.}~\bibnamefont{Wallraff}},
  \bibinfo{author}{\bibfnamefont{D.~I.} \bibnamefont{Schuster}},
  \bibinfo{author}{\bibfnamefont{A.}~\bibnamefont{Blais}},
  \bibinfo{author}{\bibfnamefont{L.}~\bibnamefont{Frunzio}},
  \bibinfo{author}{\bibfnamefont{R.-S.} \bibnamefont{Huang}},
  \bibinfo{author}{\bibfnamefont{J.}~\bibnamefont{Majer}},
  \bibinfo{author}{\bibfnamefont{S.}~\bibnamefont{Kumar}},
  \bibinfo{author}{\bibfnamefont{S.~M.} \bibnamefont{Girvin}},
  \bibnamefont{and} \bibinfo{author}{\bibfnamefont{R.~J.}
  \bibnamefont{Schoelkopf}}, \bibinfo{journal}{Nature}
  \textbf{\bibinfo{volume}{431}}, \bibinfo{pages}{162} (\bibinfo{year}{2004}).

\bibitem{leib:10}
\bibinfo{note}{M. Leib and M. J. Hartmann, arXiv:1006.2935 (2010).}

\bibitem[{\citenamefont{Hennessy et~al.}(2007)\citenamefont{Hennessy, Badolato,
  Winger, Gerace, Atature, Gulde, Falt, Hu, and Imamo\v{g}lu}}]{hennessy:07}
\bibinfo{author}{\bibfnamefont{K.}~\bibnamefont{Hennessy}},
  \bibinfo{author}{\bibfnamefont{A.}~\bibnamefont{Badolato}},
  \bibinfo{author}{\bibfnamefont{M.}~\bibnamefont{Winger}},
  \bibinfo{author}{\bibfnamefont{D.}~\bibnamefont{Gerace}},
  \bibinfo{author}{\bibfnamefont{M.}~\bibnamefont{Atature}},
  \bibinfo{author}{\bibfnamefont{S.}~\bibnamefont{Gulde}},
  \bibinfo{author}{\bibfnamefont{S.}~\bibnamefont{Falt}},
  \bibinfo{author}{\bibfnamefont{E.~L.} \bibnamefont{Hu}}, \bibnamefont{and}
  \bibinfo{author}{\bibfnamefont{A.}~\bibnamefont{Imamo\v{g}lu}},
  \bibinfo{journal}{Nature} \textbf{\bibinfo{volume}{445}},
  \bibinfo{pages}{896} (\bibinfo{year}{2007}).

\bibitem[{\citenamefont{Trupke et~al.}(2007)\citenamefont{Trupke, Goldwin,
  Darqui\'e, Dutier, Eriksson, Ashmore, and Hinds}}]{trupke:07}
\bibinfo{author}{\bibfnamefont{M.}~\bibnamefont{Trupke}},
  \bibinfo{author}{\bibfnamefont{J.}~\bibnamefont{Goldwin}},
  \bibinfo{author}{\bibfnamefont{B.}~\bibnamefont{Darqui\'e}},
  \bibinfo{author}{\bibfnamefont{G.}~\bibnamefont{Dutier}},
  \bibinfo{author}{\bibfnamefont{S.}~\bibnamefont{Eriksson}},
  \bibinfo{author}{\bibfnamefont{J.}~\bibnamefont{Ashmore}}, \bibnamefont{and}
  \bibinfo{author}{\bibfnamefont{E.~A.} \bibnamefont{Hinds}},
  \bibinfo{journal}{Phys. Rev. Lett.} \textbf{\bibinfo{volume}{99}},
  \bibinfo{pages}{063601} (\bibinfo{year}{2007}).

\bibitem[{\citenamefont{Bajcsy et~al.}(2009)\citenamefont{Bajcsy, Hofferberth,
  Balic, Peyronel, Hafezi, Zibrov, Vuletic, and Lukin}}]{bajcsy:09}
\bibinfo{author}{\bibfnamefont{M.}~\bibnamefont{Bajcsy}},
  \bibinfo{author}{\bibfnamefont{S.}~\bibnamefont{Hofferberth}},
  \bibinfo{author}{\bibfnamefont{V.}~\bibnamefont{Balic}},
  \bibinfo{author}{\bibfnamefont{T.}~\bibnamefont{Peyronel}},
  \bibinfo{author}{\bibfnamefont{M.}~\bibnamefont{Hafezi}},
  \bibinfo{author}{\bibfnamefont{A.~S.} \bibnamefont{Zibrov}},
  \bibinfo{author}{\bibfnamefont{V.}~\bibnamefont{Vuletic}}, \bibnamefont{and}
  \bibinfo{author}{\bibfnamefont{M.~D.} \bibnamefont{Lukin}},
  \bibinfo{journal}{Phys. Rev. Lett.} \textbf{\bibinfo{volume}{102}},
  \bibinfo{pages}{203902} (\bibinfo{year}{2009}).

\bibitem[{vet()}]{vetsch:09}
\bibinfo{note}{E. Vetsch and D. Reitz and G. Sagu\'{e} and R. Schmidt and S. T.
  Dawkins and A. Rauschenbeutel, arXiv:0912.1179 (2009).}

\bibitem[{\citenamefont{D\"urr et~al.}(2009)\citenamefont{D\"urr,
  Garc\'ia-Ripoll, Syassen, Bauer, Lettner, Cirac, and Rempe}}]{duerr:09}
\bibinfo{author}{\bibfnamefont{S.}~\bibnamefont{D\"urr}},
  \bibinfo{author}{\bibfnamefont{J.~J.} \bibnamefont{Garc\'ia-Ripoll}},
  \bibinfo{author}{\bibfnamefont{N.}~\bibnamefont{Syassen}},
  \bibinfo{author}{\bibfnamefont{D.~M.} \bibnamefont{Bauer}},
  \bibinfo{author}{\bibfnamefont{M.}~\bibnamefont{Lettner}},
  \bibinfo{author}{\bibfnamefont{J.~I.} \bibnamefont{Cirac}}, \bibnamefont{and}
  \bibinfo{author}{\bibfnamefont{G.}~\bibnamefont{Rempe}},
  \bibinfo{journal}{Phys. Rev. A} \textbf{\bibinfo{volume}{79}},
  \bibinfo{pages}{023614} (\bibinfo{year}{2009}).

\bibitem[{\citenamefont{Syassen et~al.}(2008)\citenamefont{Syassen, Bauer,
  Lettner, Volz, Dietze, Garc\'ia-Ripoll, Cirac, Rempe, and
  D\"urr}}]{syassen:08}
\bibinfo{author}{\bibfnamefont{N.}~\bibnamefont{Syassen}},
  \bibinfo{author}{\bibfnamefont{D.~M.} \bibnamefont{Bauer}},
  \bibinfo{author}{\bibfnamefont{M.}~\bibnamefont{Lettner}},
  \bibinfo{author}{\bibfnamefont{T.}~\bibnamefont{Volz}},
  \bibinfo{author}{\bibfnamefont{D.}~\bibnamefont{Dietze}},
  \bibinfo{author}{\bibfnamefont{J.~J.} \bibnamefont{Garc\'ia-Ripoll}},
  \bibinfo{author}{\bibfnamefont{J.~I.} \bibnamefont{Cirac}},
  \bibinfo{author}{\bibfnamefont{G.}~\bibnamefont{Rempe}}, \bibnamefont{and}
  \bibinfo{author}{\bibfnamefont{S.}~\bibnamefont{D\"urr}},
  \bibinfo{journal}{Science} \textbf{\bibinfo{volume}{320}},
  \bibinfo{pages}{1329} (\bibinfo{year}{2008}).

\bibitem[{\citenamefont{Breuer and Petruccione}(2006)}]{breuer:os}
\bibinfo{author}{\bibfnamefont{H.-P.} \bibnamefont{Breuer}} \bibnamefont{and}
  \bibinfo{author}{\bibfnamefont{F.}~\bibnamefont{Petruccione}},
  \emph{\bibinfo{title}{The Theory of Open Quantum Systems}}
  (\bibinfo{publisher}{Oxford University Press}, \bibinfo{address}{Oxford},
  \bibinfo{year}{2006}).

\bibitem[{\citenamefont{Liu et~al.}(2001{\natexlab{b}})\citenamefont{Liu,
  Dutton, Behroozi, and Hau}}]{liu:01}
\bibinfo{author}{\bibfnamefont{C.}~\bibnamefont{Liu}},
  \bibinfo{author}{\bibfnamefont{Z.}~\bibnamefont{Dutton}},
  \bibinfo{author}{\bibfnamefont{C.~H.} \bibnamefont{Behroozi}},
  \bibnamefont{and} \bibinfo{author}{\bibfnamefont{L.~V.} \bibnamefont{Hau}},
  \bibinfo{journal}{Nature} \textbf{\bibinfo{volume}{409}},
  \bibinfo{pages}{490} (\bibinfo{year}{2001}{\natexlab{b}}).

\bibitem[{\citenamefont{Lieb and Liniger}(1963)}]{lieb:63}
\bibinfo{author}{\bibfnamefont{E.~H.} \bibnamefont{Lieb}} \bibnamefont{and}
  \bibinfo{author}{\bibfnamefont{W.}~\bibnamefont{Liniger}},
  \bibinfo{journal}{Phys. Rev.} \textbf{\bibinfo{volume}{130}},
  \bibinfo{pages}{1605} (\bibinfo{year}{1963}).

\bibitem[{\citenamefont{Girardeau}(1960)}]{girardeau:60}
\bibinfo{author}{\bibfnamefont{M.}~\bibnamefont{Girardeau}},
  \bibinfo{journal}{J. Math. Phys.} \textbf{\bibinfo{volume}{1}},
  \bibinfo{pages}{516} (\bibinfo{year}{1960}).

\bibitem[{\citenamefont{Astrakharchik et~al.}(2005)\citenamefont{Astrakharchik,
  Boronat, Casulleras, and Giorgini}}]{astrakharchik:05}
\bibinfo{author}{\bibfnamefont{G.~E.} \bibnamefont{Astrakharchik}},
  \bibinfo{author}{\bibfnamefont{J.}~\bibnamefont{Boronat}},
  \bibinfo{author}{\bibfnamefont{J.}~\bibnamefont{Casulleras}},
  \bibnamefont{and} \bibinfo{author}{\bibfnamefont{S.}~\bibnamefont{Giorgini}},
  \bibinfo{journal}{Phys. Rev. Lett.} \textbf{\bibinfo{volume}{95}},
  \bibinfo{pages}{190407} (\bibinfo{year}{2005}).

\bibitem[{\citenamefont{Batchelor et~al.}(2005)\citenamefont{Batchelor, Bortz,
  Guan, and Oelkers}}]{batchelor:05}
\bibinfo{author}{\bibfnamefont{M.~T.} \bibnamefont{Batchelor}},
  \bibinfo{author}{\bibfnamefont{M.}~\bibnamefont{Bortz}},
  \bibinfo{author}{\bibfnamefont{X.~W.} \bibnamefont{Guan}}, \bibnamefont{and}
  \bibinfo{author}{\bibfnamefont{N.}~\bibnamefont{Oelkers}},
  \bibinfo{journal}{J. Stat. Mech.} \textbf{\bibinfo{volume}{L}},
  \bibinfo{pages}{10001} (\bibinfo{year}{2005}).

\bibitem[{\citenamefont{Haller et~al.}(2009)\citenamefont{Haller, Gustavsson,
  Mark, Danzl, Hart, Pupillo, and N\"agerl}}]{haller:09}
\bibinfo{author}{\bibfnamefont{E.}~\bibnamefont{Haller}},
  \bibinfo{author}{\bibfnamefont{M.}~\bibnamefont{Gustavsson}},
  \bibinfo{author}{\bibfnamefont{M.~J.} \bibnamefont{Mark}},
  \bibinfo{author}{\bibfnamefont{J.~G.} \bibnamefont{Danzl}},
  \bibinfo{author}{\bibfnamefont{R.}~\bibnamefont{Hart}},
  \bibinfo{author}{\bibfnamefont{G.}~\bibnamefont{Pupillo}}, \bibnamefont{and}
  \bibinfo{author}{\bibfnamefont{H.-C.} \bibnamefont{N\"agerl}},
  \bibinfo{journal}{Science} \textbf{\bibinfo{volume}{325}},
  \bibinfo{pages}{1224} (\bibinfo{year}{2009}).

\bibitem[{com({\natexlab{a}})}]{comment2}
\bibinfo{note}{Note that fast oscillating spin coherences may still be produced
  via the $\ket{2}\leftrightarrow\ket{4}$ transition, but this process is
  off-resonant and thus suppressed (see Appendix~\ref{bath}).}

\bibitem[{com({\natexlab{b}})}]{comment1}
\bibinfo{note}{C. Cohen-Tannoudji, J. Dupont-Roc, G. Grynberg,
  \textit{Atom-Photon Interactions} (Wiley, New York, 1992), Sec. III.C.3.}

\bibitem[{\citenamefont{Scully and Zubairy}(1997)}]{zubairy:qo}
\bibinfo{author}{\bibfnamefont{M.~O.} \bibnamefont{Scully}} \bibnamefont{and}
  \bibinfo{author}{\bibfnamefont{M.~S.} \bibnamefont{Zubairy}},
  \emph{\bibinfo{title}{Quantum Optics}} (\bibinfo{publisher}{Cambridge
  University Press}, \bibinfo{address}{Cambridge}, \bibinfo{year}{1997}).

\bibitem[{\citenamefont{Kiffner and Dey}(2009)}]{kiffner:09}
\bibinfo{author}{\bibfnamefont{M.}~\bibnamefont{Kiffner}} \bibnamefont{and}
  \bibinfo{author}{\bibfnamefont{T.~N.} \bibnamefont{Dey}},
  \bibinfo{journal}{Phys. Rev. A} \textbf{\bibinfo{volume}{79}},
  \bibinfo{pages}{023829} (\bibinfo{year}{2009}).

\end{thebibliography}
\end{document}